\documentclass[useAMS,usenatbib, epsfig]{mn2e}
\usepackage{amssymb}
\usepackage{amsmath}
\usepackage{amstext}
\usepackage{deluxetable}
\usepackage{epsfig}
\usepackage{thumbpdf}
\usepackage{epstopdf}
\usepackage{graphicx}
\usepackage[english]{babel}
\usepackage[utf8]{inputenc}
\usepackage{natbib}
\usepackage{xargs}                   
\usepackage[pdftex,dvipsnames]{xcolor} 
 \usepackage[colorinlistoftodos,prependcaption,textsize=tiny]{todonotes}
\usepackage[colorinlistoftodos,prependcaption,textsize=tiny]{todonotes}
\newcommandx{\unsure}[2][1=]{\todo[linecolor=red,backgroundcolor=red!25,bordercolor=red,#1]{#2}} 
\newcommandx{\change}[2][1=]{\todo[linecolor=blue,backgroundcolor=blue!25,bordercolor=blue,#1]{#2}} 
\newcommandx{\info}[2][1=]{\todo[linecolor=OliveGreen,backgroundcolor=OliveGreen!25,bordercolor=OliveGreen,#1]{#2}} 
\newcommandx{\improvement}[2][1=]{\todo[linecolor=Plum,backgroundcolor=Plum!25,bordercolor=Plum,#1]{#2}}
\newcommandx{\thiswillnotshow}[2][1=]{\todo[disable,#1]{#2}}
\usepackage{lmacs}
\usepackage{savesym}

\usepackage[citecolor=blue]{hyperref}
\hypersetup{colorlinks=true}
\newcommand{\avg}[1]{\left< #1 \right>} 
 \usepackage{pdf14}

\def\Zsun{${\rm Z}_{\odot}$}
\def\Msun{${\rm M}_{\odot}$}
\newcommand{\Hbd}{\rm{H}$\beta$}
\newcommand{\Had}{\rm{H}$\alpha$}
\def\EBV{E($B-V$)}
\def\Rv{$R_V$}
\def\OIIId{[{\sc O III}] $\lambda$5007}
\def\NIId{[{\sc N II}] $\lambda$6583}
\def\OIIIud{[{\sc O III}] $\lambda$$\lambda$4959, 5007}
\def\OIIIt{[{\sc O III}] $\lambda$4363}
\def\OIIIHb{[{\sc O   III}]/H$\beta$}
\def\NIIHa{[{\sc N II}]/H$\alpha$}

\DeclareRobustCommand{\ion}[2]{%
\relax\ifmmode
\ifx\testbx\f@series
{\mathbf{#1\,\mathsc{#2}}}\else
{\mathrm{#1\,\mathsc{#2}}}\fi
\else\textup{#1\,{\mdseries\textsc{#2}}}%
\fi}

\title[SN environments with MUSE]{Characterising the environments of supernovae with MUSE}

\author[Galbany et al.]{L. Galbany$^{1,2}$\thanks{E-mail: lgalbany@das.uchile.cl}, 
J. P. Anderson$^{3}$,  
F. F Rosales-Ortega$^{4}$,
H. Kuncarayakti$^{1,2}$,
\newauthor  
T. Kr\"{u}hler$^{3,5}$, 
S. F. S\'anchez$^{6}$, 
J. Falc\'on-Barroso$^{7}$, 
E. P\'erez$^{8}$,
J. C. Maureira$^{9}$, 
\newauthor 
M. Hamuy$^{2,1}$, 
S. Gonz\'alez-Gait\'an$^{1,2}$, 
F. F\"orster$^{1,9}$, 
V. Moral$^{2,9}$\\
$^{1}$Millennium Institute of Astrophysics, Universidad de Chile, Casilla 36-D, Santiago, Chile.\\
$^{2}$Departamento de Astronom\'ia, Universidad de Chile, Casilla 36-D, Santiago, Chile.\\
$^{3}$European Southern Observatory, Alonso de Cordova 3107 Casilla 19001 - Vitacura -Santiago, Chile.\\
$^{4}$Instituto Nacional de Astrof\'isica,, \'Optica y Electr\'onica, Luis E. Erro 1, 72840 Tonantzintla, Puebla, Mexico.\\
$^{5}$Max-Planck-Institut f\"{u}r extraterrestrische Physik, Giessenbachstra\ss e, 85748 Garching, Germany\\
$^{6}$Instituto de Astronom\'ia, Universidad Nacional Aut\'onoma de M\'exico, A.P. 70-264, 04510, M\'exico DF, Mexico.\\
$^{7}$Instituto de Astrof\'isica de Canarias, E-38205 La Laguna, Tenerife, Spain.\\
$^{8}$Instituto de Astrof\'isica de Andaluc\'ia (CSIC), Glorieta de la Astronom\'ia s/n, Aptdo. 3004, E18080-Granada, Spain.\\
$^{9}$Centro de Modelamiento Matem\'atico, Universidad de Chile, Av. Blanco Encalada 2120 Piso 7, Santiago, Chile.}
\date{Received date: 9 October 2015; accepted date: ***}

\begin{document}
\maketitle

\begin{abstract}
We present a statistical analysis of the environments of 11 supernovae (SNe) which occurred in 6 nearby galaxies (z $\lesssim$ 0.016). 
All galaxies were observed with MUSE, the high spatial resolution integral field spectrograph mounted to the 8m VLT UT4. 
These data enable us to map the full spatial extent of host galaxies up to $\sim$3 effective radii. 
In this way, not only can one characterise the specific host environment of each SN, one can compare their properties with stellar populations within the full range of other environments within the host. 
We present a method that consists of selecting all {\sc Hii} regions found within host galaxies from 2D extinction-corrected H$\alpha$ emission maps. 
These regions are then characterised in terms of their H$\alpha$ equivalent widths, star formation rates, and oxygen abundances. 
Identifying {\sc Hii} regions spatially coincident with SN explosion sites, we are thus able to determine where within the distributions of host galaxy e.g. metallicities and ages each SN is found, thus providing new constraints on SN progenitor properties. 
This initial pilot study using MUSE opens the way for a revolution in SN environment studies where we are now able to study multiple environment SN progenitor dependencies using a single instrument and single pointing.
\end{abstract}



\section{Introduction}

Supernovae (SNe) play a key role in our understanding of stellar and galaxy evolution. However, despite their importance the exact physical explosion mechanism and the nature of the progenitor stars of each SN type are not yet completely constrained.
Core collapse (CC) SNe are the final stage in the evolution of massive stars ($\gtrsim$8 M$\odot$), while type Ia SNe are the result of thermonuclear explosions of carbon-oxygen white dwarfs.
Both types show a wide diversity in their photometric and spectroscopic properties, which can be related to different progenitor star characteristics \citep{2003ApJ...582..905H,2005A&A...443..649M, 2014NewAR..62...15R, 2015MNRAS.451.2212G}.

One possible approach to constrain progenitor systems is to study the environments where SNe explode. 
Differences in the ages and/or metallicities of the progenitor stars can explain the differences in observed SN properties.
Previous works have tried this approach:  
\cite{2008ApJ...673..999P} and \cite{2012ApJ...759..107K} used SDSS spectra centered at the galaxy core (although in a few cases at other locations), and found an increasing ratio of number of SN Ibc with respect SN II as the metallicity increases. 
Also \cite{2014ApJ...791...57S} claimed differences in galaxy stellar population average age between SN Ia and CC SN hosts.
Further works tried to characterize the local properties of the SN environment within host galaxies. 
\cite{2010MNRAS.407.2660A} and \cite{2011A&A...530A..95L} found no statistically significant difference between the local metallicity of SNe Ib and Ic, obtaining slit or fiber spectra at SNe positions. The lack of difference was reinforced by \cite{2012ApJ...758..132S} studying a survey of SNe from untargeted surveys. 
However, \cite{2011ApJ...731L...4M} concluded that significant differences do exist between environment metallicities of SNe Ib and SNe Ic. Using photometric measurements, \cite{2008ApJ...687.1201K} found that SNe Ic in general explode in the brightest regions of their hosts when compared to other SN types. Meanwhile, \cite{2012MNRAS.424.1372A} found a sequence of increasing association of SNe to host galaxy on-going star-formation (as traced by H$\alpha$ emission), and concluded that this implied an increasing progenitor mass sequence from SNe II to SNe Ib and finally SNe Ic arising from the most massive progenitors.

In the case of SNe Ia, several authors have attempted to use environment information to a) further understand progenitor systems, and b) reduce residuals in Hubble diagrams. These have involved using the galactocentric distances as environment property proxies \citep{2012ApJ...755..125G, 2012MNRAS.424.1372A,2015MNRAS.448..732A}, using photometric measurements at explosion sites \citep{2015MNRAS.448..732A,2015Sci...347.1459K}, together with environment spectroscopy \citep{2013A&A...560A..66R,2015A&A...580A.131T}, or by indirect approximations of the central metallicity and applying decreasing gradients \citep{2009A&A...503..137B}.

While the above studies used distinct observations and techniques, they all had one common theme. They either used spectroscopy at a specific location within host galaxies, i.e. the SN explosion site, in isolation from all other stellar populations found within hosts, or they used photometric measurements of host galaxies. 
In the former one obtains detailed spectroscopic information but lacks spatial information, while in the latter one has spatial but no spectral information. 
In addition, each technique used a specific observation or set of observations to investigate one individual property, be it metallicity (e.g. through emission line spectroscopy) or age (through association of SNe to star-forming regions). The introduction of wide-field Integral Field Spectroscopy (IFS) allows --for the first time-- simultaneous investigation of both spatial and spectral information, hence enabling conclusions to be made on SN progenitor ages and metallicities, from a range of techniques from a single telescope pointing. This is the direction of the current paper: to show how wide-field and high spatial resolution IFS can be used to study SN environment and take this field in new directions.

\cite{2013AJ....146...30K, 2013AJ....146...31K} used Integral Field Spectroscopy (IFS) observations of high spatial resolution (but with small field of view, FoV) to disentangle single stellar clusters where the progenitor star could have been born, and compared the masses and metallicities for different SN progenitor parent stellar clusters.
Similar studies have been performed with galaxies observed by the CALIFA survey using PPAK/PMAS, an Integral Field Unit (IFU) with large FoV and a spatial resolution of 1 arcsec$^2$. 
\cite{2014A&A...572A..38G} have compiled a large sample of 128 SNe and probed for the first time spectroscopically the different association of different SN types to the star-formation of their environment. 
Moreover, the differences in the local environmental oxygen abundance for different SN types have been also investigated: a sequence from SN Ia occurring on average in metal-rich environments, through SNe IIn, Ic, II, Ib, and IIb was found (Galbany et al. in prep.).
A recent review of SN environmental studies can be found in \cite{2015PASA...32...19A}, which concentrates on the more traditional techniques described above, but also presents some preliminary analysis of data used in the current work and speculates as to future directions of environmental studies.

In this paper, we aim to probe the capabilities of the Multi-Unit Spectroscopic Explorer (MUSE) IFU recently mounted to the 8m VLT UT4, by obtaining wide-field and high-spatial resolution spectroscopy of the whole extent of galaxies that have hosted several detected supernovae, in order to compare SN parent stellar populations with both global and positionally distinct regions of the host galaxy.
Our sample consists of 6 galaxies which hosted a total of 11 SNe. 
MUSE allows for the first time to obtain observations of the full spatial extent of a single SN host galaxy in both a small number of pointings (2 at most in our case), together with unprecedented spatial resolution. This allows a spectroscopic study of multiple distinct regions within the galaxy of locations which have/have not hosted SNe.
Using these data we will investigate whether SNe explode in higher/lower metallicity/age/extinction populations than the overall galaxy, and use these parameters to further probe the nature of SNe progenitor populations in the context of the overall host galaxy.

This work will serve as a pilot study which will enable the community to further define how MUSE can be used in this field of SN galaxy environment studies. 
Recently, the All-weather MUSE Integral-field Nearby Galaxies (AMUSING, Anderson et al. in prep.) survey has started compiling a large number of nearby SN host galaxies that will serve as a statistical sample to study the environments of SNe.
In this first paper using MUSE to study SN environments, we investigate the different avenues opened by such an instrument and present techniques which can be used in future studies to extract measurements from the deluge of data now available for SN environment studies.

This paper is organized as follows: In section \S\ref{sec:obs}, we describe the characteristics of the observations and the compilation of the sample. Then, in section \S\ref{sec:ana} we describe the analysis performed to the data to extract the needed spectral parameters. In section \S\ref{sec:res} the results for the 10 SNe are presented, and finally in section \S\ref{sec:disc} we summarize and give our conclusions.


\begin{table*}\scriptsize
\caption{Description of the observations}           
\label{tab:obs}    
\centering        
\begin{tabular}{lccccc}
\hline\hline   
Galaxy name  & Observations  (2014)        & ESO project ID &  PI & Exp. time [s]  & Seeing ["]\\
\hline
NGC 2906      & COMMISSIONING July 28th & 60.A-9100(B) &    	MUSE TEAM &   4 $\times$ 300  & 0.88 \\
ESO 362-18   & COMMISSIONING July 28th &60.A-9100(A) &  	MUSE TEAM & 5 $\times$ 1200  & 0.98 \\
NGC 6754      & SV June  28th and 30th & 60.A-9329(A)      &  Galbany &  3 $\times$ 940 (E), 3 $\times$ 940 (W)  & 0.95 (E), 2.65 (W)\\
NGC 7742      &  SV June  23th & 60.A-9301(A)      &  Sarzi & 2 $\times$ 1800   & 1.14 \\
IC 1158           & SV June  24th &  60.A-9319(A)    & Mu\~noz-Mateos &  2 $\times$ 540 (in), 2 $\times$ 910 (out) & 0.98 (in), 1.06 (out) \\
NGC 7469      &  SV August 19th        & 60.A-9339(A)      &  Hawthorn/Marconi/Salvato &  4 $\times$ 600  & 1.23 \\
 \hline
\end{tabular}
\end{table*}
\begin{table*}\scriptsize
\caption{Summary of the galaxy and SN sample. Galaxies ordered by Right Ascension (RA).}           
\label{tab:sample}    
\centering        
\begin{tabular}{lccclcclccc}
\hline\hline   
Galaxy name    &  RA 		&  DEC        & redshift     & Morphology & 	$E(B-V)$ & SN name& SN type	  & Off. EW & Off. NS & Separation	\\ 
    &   		&         &     &  & [mag] & &   & ["] & ["] & [']\\ 
\hline
ESO 362-     18 & 05 19 35.810 & --32 39 28.02 & 0.012445  & SB0/a?(s) pec &0.017   &2010jr       &II              &  --16.8 &  +13.6 & 0.351  \\
NGC 2906         & 09 32 06.247 &+08 26 30.86  & 0.007138  & Scd?               &  0.047  &2005ip     &IIn              &  +02.8  & +14.2  & 0.239  \\
IC 1158              &  16 01 34.072 &+01 42 28.19    & 0.006428 & SAB(r)c?         &0.114  &2000cb      &II  87A-like               &--29.5  & --04.2   &0.490 \\
NGC 6754         &  19 11 25.752 & --50 38 31.96& 0.010864 & SAB(rs)bc       &0.070   &2000do    &Ia              & +04.5  & +07.6 & 0.125  \\
                         &                      &                    &                 &                         &            & 2005cu   &II               & --12.0  & +03.0 & 0.194\\
                         &                      &                    &                 &                        &            &1998dq      &Ia             &  --18.7  & +05.0 & 0.319\\
                         &                      &                    &                 &                        &            &1998X        &II              &-23.8  & --03.1  &0.405\\
NGC 7469         & 23 03 15.674 &+08 52 25.28 & 0.016317 & (R')SAB(rs)a    &0.069  &2000ft       & II radio          &--01.7  & +00.0  &0.031\\
                         &                     &                     &                 &                         &           &2008ec     &Ia              & +13.7  & --07.4   &0.256\\
NGC 7742         &  23 44 15.710 & +10 46 01.55  & 0.005547 & SA(r)b              &0.055  &1993R       &Ia 91bg-like  &+08.0  & +05.0  &0.218  \\
                         &                     &                     &                 &                         &           &2014cy     &II              & +01.0  & +12.0   &0.198\\
 \hline
\end{tabular}
\end{table*}

\section{Observations} \label{sec:obs}

MUSE \citep{2014Msngr.157...13B} is a new IFS located at the Nasmyth B focus of Yepun, the VLT UT4 telescope at Cerro Paranal Observatory.
It has a modular structure composed of 24 identical IFU modules that together sample, in Wide Field Mode (WFM), a near-contiguous 1 squared arcmin FoV. 
The instrument provides a FoV of 1 arcmin$^2$ with spaxels of 0.2"$\times$0.2", and a wavelength coverage of 4650-9300~\AA~with a mean resolution of R$\sim$3000.
The combination of a large FoV with a spatial sampling sufficient to properly sample the PSF even under good seeing conditions makes MUSE a unique instrument for a variety of science applications. Although the main motivation for developing this new instrument is the observation of faint galaxies at medium to high redshift, some very promising applications exist for the investigation of nearby galaxies.

NGC 6754 was observed in the context of Programme 60.A-9329 (PI Galbany) of the first MUSE Science Verification run. 
All data from the MUSE commissioning and SV were made publicly available immediately after the observations.
After a thorough search for SNe in other public targets, a sample of 6 galaxies that hosted 11 SNe  (4 Ia, 7 II) have been compiled.
Besides 7 typical SN II and Ia, the following four SNe had more precise classifications:
SN 2000ft was discovered in NGC 7469 by using radio observations and showed similar evolution properties than other compact radio sources identified as SN II \citep{2006ApJ...638..938A}; 
SN 2000cb showed similar photometric behaviour than SN 1987A, which are thought to be SN II resulting from the explosion of blue supergiants \citep{2005MNRAS.360..950P, 2012A&A...537A.140T};
SN 1993R was typed as a peculiar SN similar to the subluminous SN Ia class 1991bg-like, but with stronger Ca II NIR triplet \citep{1993IAUC.5842....2F}.
Finally, SN 2005ip spectra showed that it was interacting with the surrounding circumstellar medium and it was typed as SN IIn \citep{2012ApJ...756..173S}.%
The programs that proposed the observations of these 5 galaxies, observation periods, exposure times, and seeing at the time of observations are summarized in Table \ref{tab:obs}.

MUSE delivers an impressive set of $\sim$100,000 spectra per pointing covering most of the optical domain. 
For NGC 6754, the observations were divided in two pointings, one for each eastern and western part of the galaxy. 
The final cube is the result of 3 observations of 940 sec of exposure time, where the second and third were slightly shifted (2 arcsec NE and SW) in order to provide a uniform coverage of the FoV and limit systematic errors in flatfielding and sky-subtraction, and rotated 90 deg to be able to reconstruct an homogenous coverage including the edges. 
In between the two science pointings another 170 sec observation of the sky was performed to be able to remove the telluric contribution from the science images.
The reduction of the raw data was performed with {\sc REFLEX} \citep{2013A&A...559A..96F} using version 0.18.5 of the MUSE pipeline \citep{2014ASPC..485..451W} with default parameters, which consists of the standard procedures of bias subtraction, flat fielding, sky-subtraction, wavelength calibration, and flux calibration.
\cite{2015A&A...573A.105S} presented a complete census of the {\sc Hii} regions in the galaxy NGC 6754 using the same observations presented here. 
These were used to derive the radial oxygen abundance gradient and estimate the typical mixing scale length. 
Evidence of an azimuthal variation in the oxygen abundance was found, which may be connected with radial migration. 

\begin{figure*}
\centering
\includegraphics[trim=6cm 19.1cm 6cm 1.6cm, clip=true,width=0.325\textwidth]{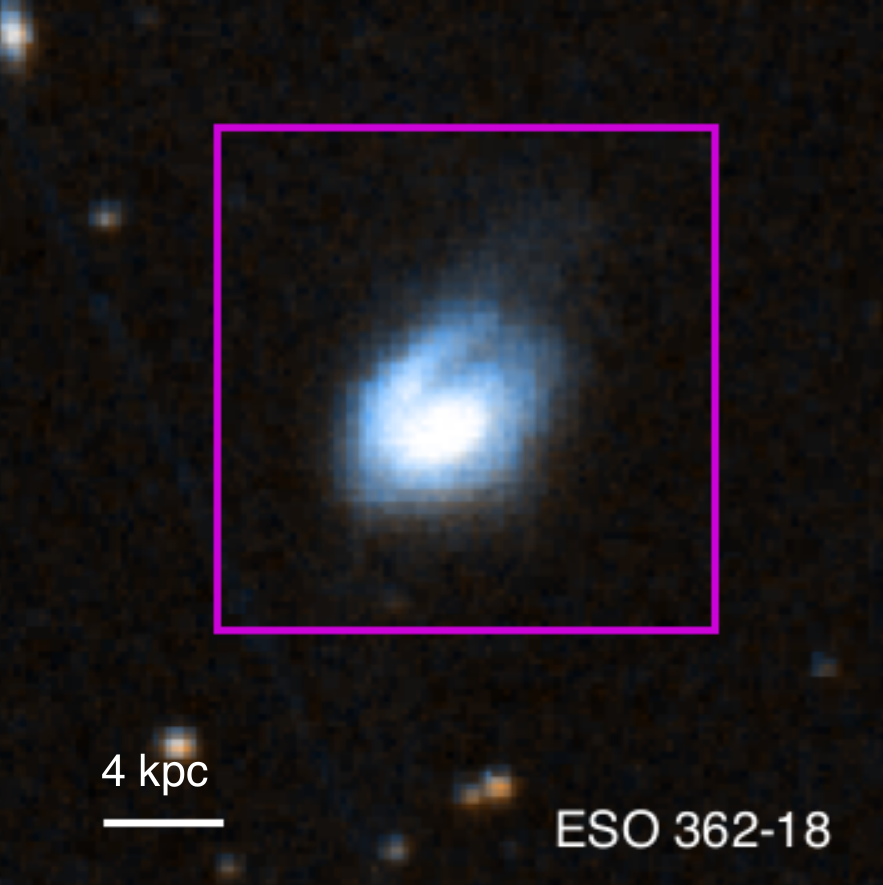}
\includegraphics[trim=2.8cm 2.8cm 2.8cm 2.8cm, clip=true,width=0.325\textwidth]{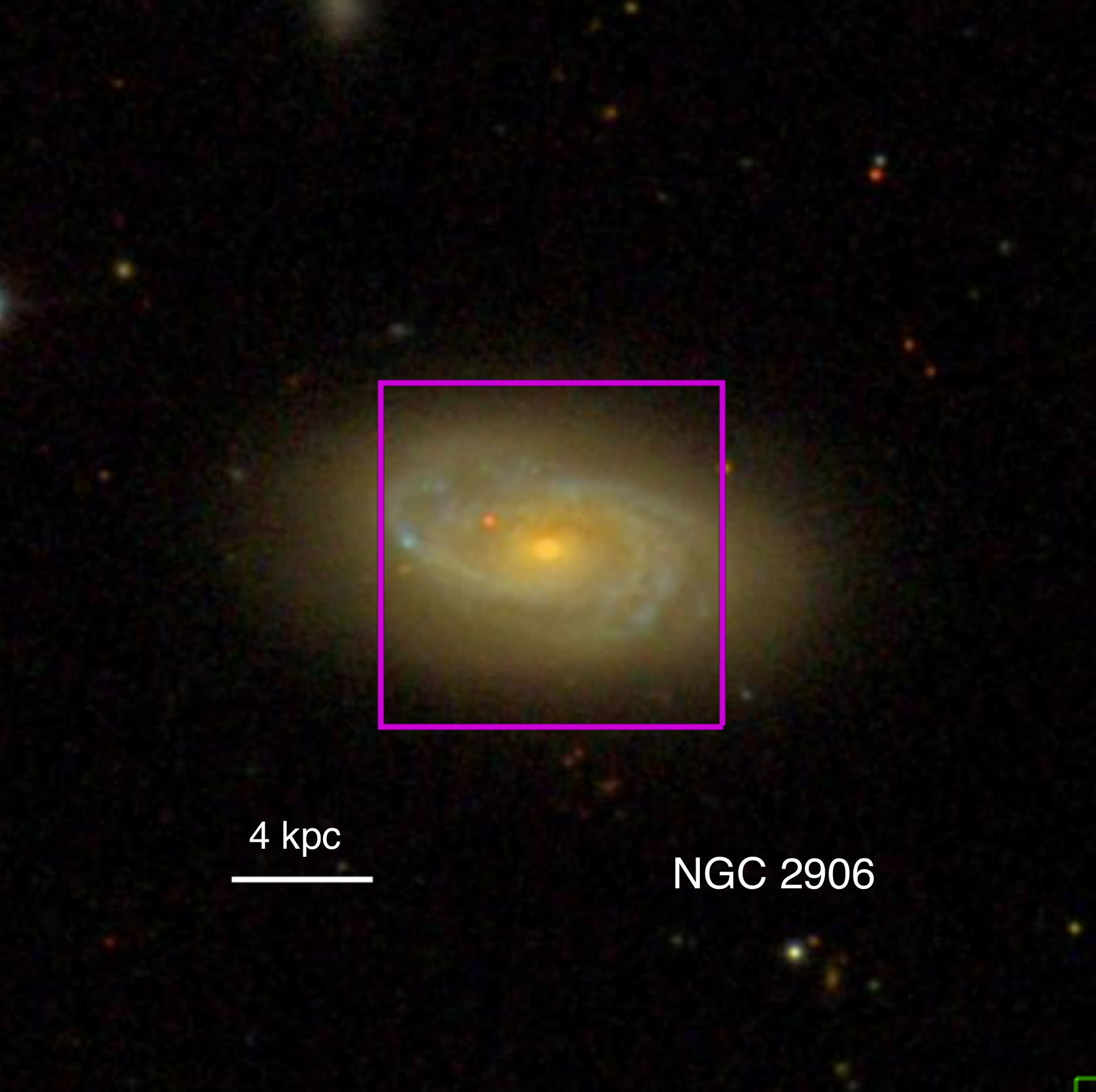}
\includegraphics[trim=2.8cm 2.8cm 2.8cm 2.8cm, clip=true,width=0.325\textwidth]{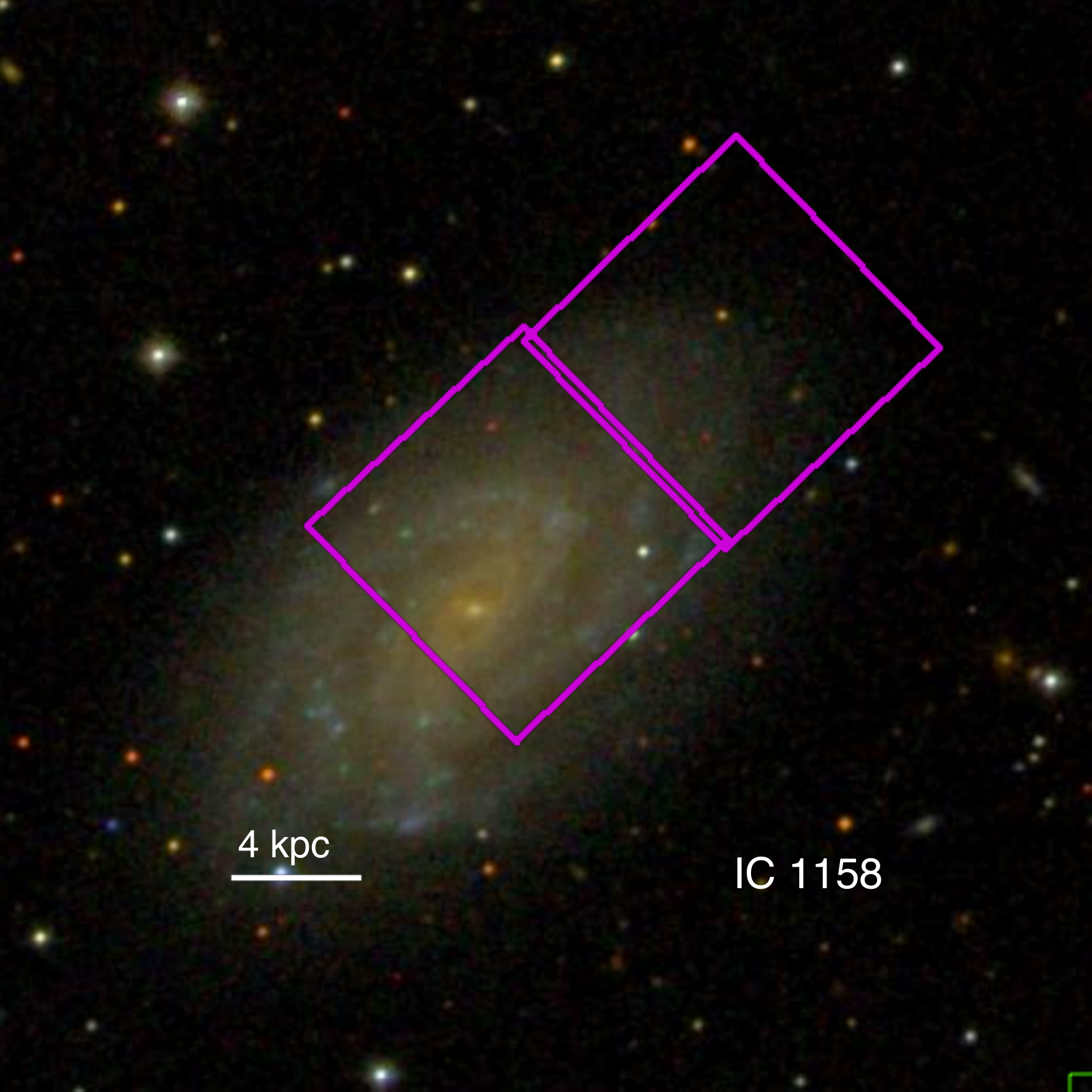}
\includegraphics[trim=6cm 19.1cm 6cm 1.6cm, clip=true,width=0.325\textwidth]{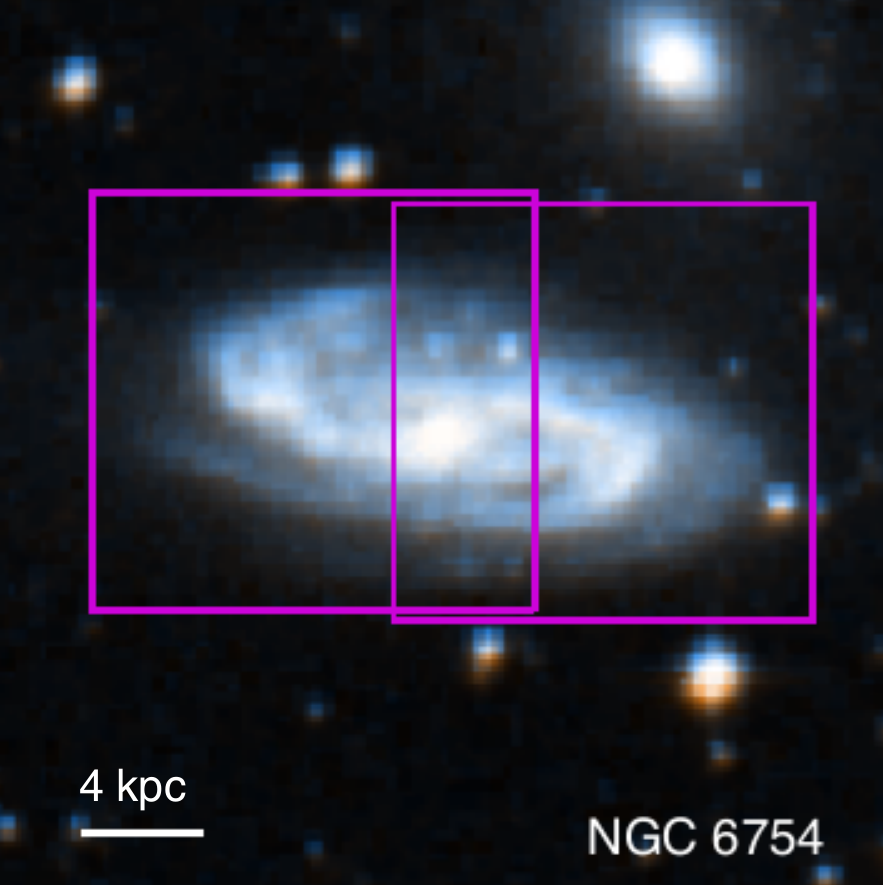}
\includegraphics[trim=2.8cm 2.8cm 2.8cm 2.85cm, clip=true,width=0.325\textwidth]{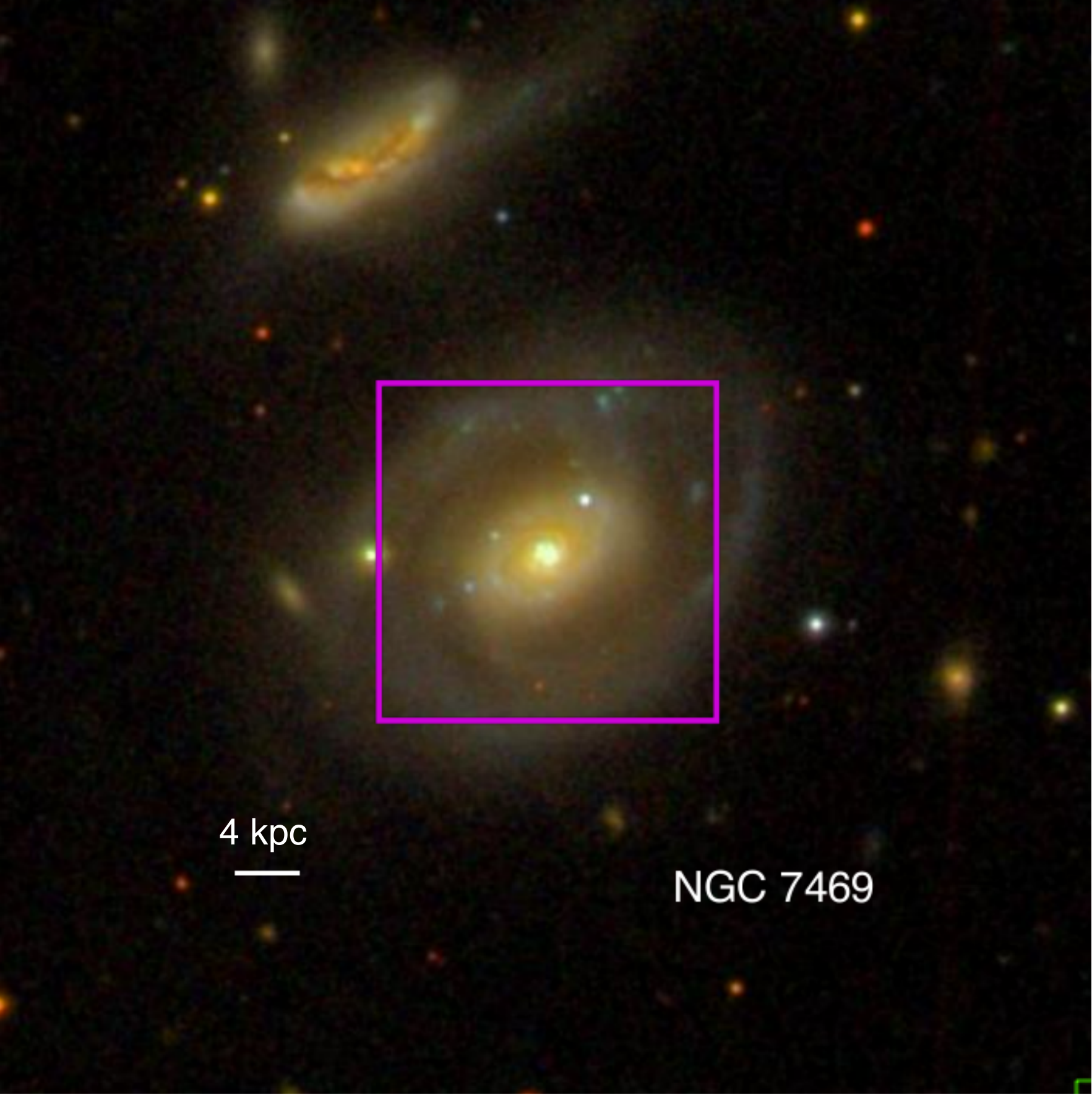}
\includegraphics[trim=2.8cm 2.8cm 2.8cm 2.9cm, clip=true,width=0.325\textwidth]{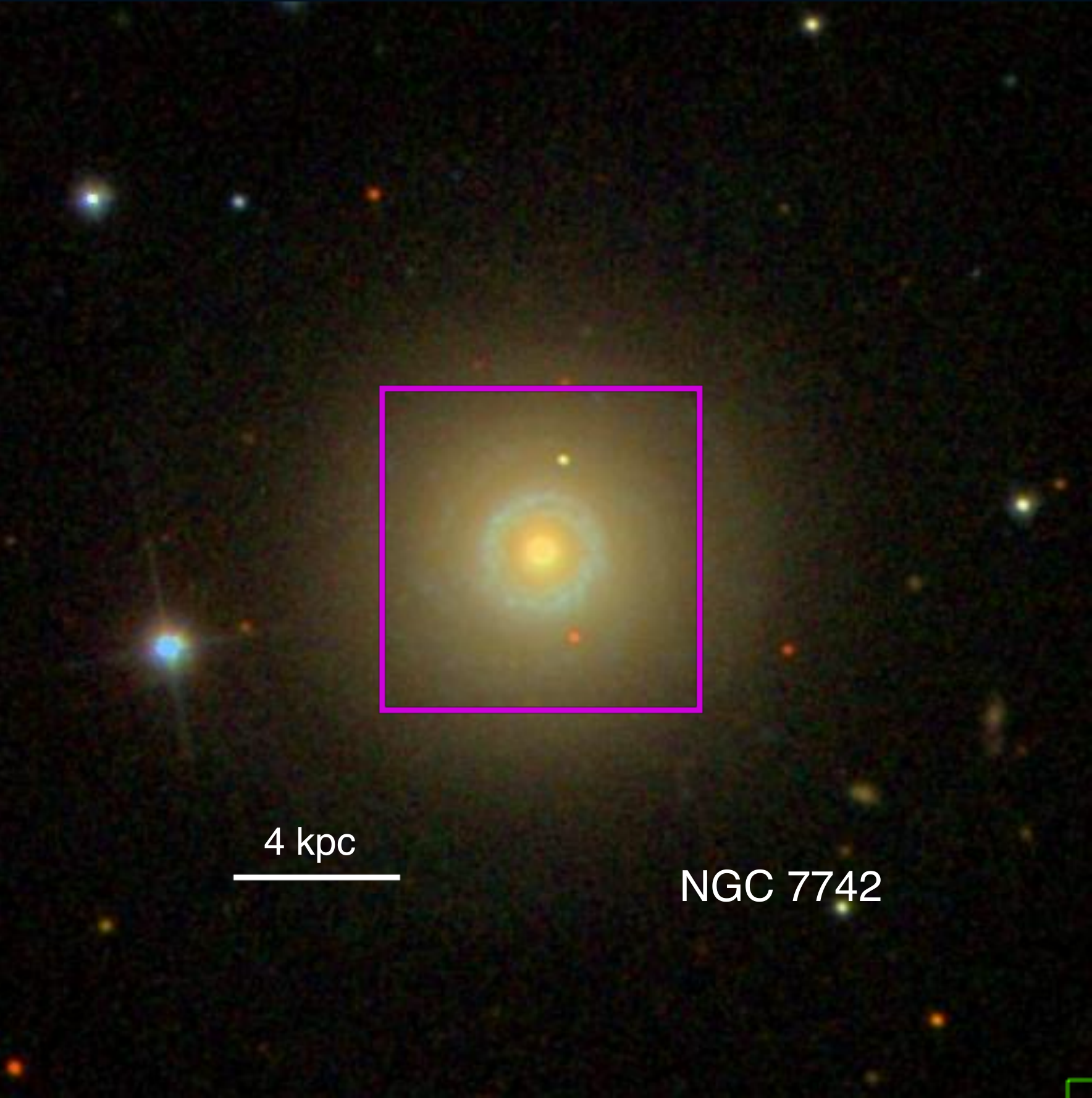}
\caption{Images of the galaxies obtained from the SDSS when available, and DSS when not. The MUSE field of view is over plotted.}
\label{fig:sdss}
\end{figure*}

For IC 1158 also 2 pointings were obtained, and the other 4 galaxies were observed with only 1 pointing. They all were reduced in the same way than NGC 6754. Details of all SN and host galaxy properties are listed
in Table 2. In Figure \ref{fig:sdss} we show the MUSE FoV over the Digitized Sky Survey (DSS) and the Sloan Digital Sky Survey (SDSS) images of the galaxies composing our sample, and a bar which gives an idea of their physical scales.

\section{Analysis}\label{sec:ana}

As noted in the introduction, we have previous experience of analysing wide-field IFS observations of SN host galaxies through the CALIFA survey (see SN environment analysis in \citealt{2012A&A...545A..58S} and \citealt{2014A&A...572A..38G}). Hence, the same techniques were adapted to MUSE data. However, we also note an important difference: in the CALIFA analysis SN explosion site stellar population parameters were compared to those of the overall host, however here we go one step further in comparing SN explosion site environments to all other stellar populations found within hosts, therefore fully utilising the MUSE observations.

A major difference between MUSE and previous large FOV IFS instruments is the impressive 0.2"$\times$0.2" pixel scale. However, for nearby galaxy studies where one is interested in extracting {\sc Hii} region parameters, this actually complicates interpretations of spectra. In such high spatial resolution spectra it is possible that each spaxel actually contains only part of an {\sc Hii} region. In addition, it has been shown that there exist significant diffuse H$\alpha$ emission within galaxies which does not arise from star-forming {\sc Hii} regions \citep{2002ApJ...565L..79C,2005A&A...438..599G,2014A&A...563A..49S}. Therefore, in order to analyse data in a consistent manner, and also to enable {\sc Hii} region assumptions to be secure when analysing spectra, we choose to focus our analysis on spectra spatially aggregated by identifying {\sc Hii} regions within host galaxies (using specific software created for this job, which is detailed below). In the next sections we first work with individual spaxel spectra in order to obtain 2D extinction-corrected H$\alpha$ emission maps in order to identify individual {\sc Hii} regions. Then, the spectral extraction process is repeated working with {\sc Hii} regions, and the parameters for our analysis are hence obtained. Later, we also perform our analysis on individual spaxel spectra, and compare the results between the two methods.

\subsection{Stellar continuum fitting}\label{sec:sf}

All the original spectra were corrected for the Milky Way dust extinction using the dust maps of \cite{2011ApJ...737..103S} and applying the standard Galactic reddening law with $R_V$ = 3.1 \citep{1999PASP..111...63F}. The spectra were then corrected to rest frame wavelengths.

Adopting the basic assumption that the star formation history of a galaxy can be approximated as the sum of discrete star formation bursts, the observed stellar spectrum of a galaxy can be represented as the sum of spectra of single stellar population (SSP) with different ages and metallicities. 
We used the {\sc STARLIGHT} code \citep{2009RMxAC..35..127C, 2005MNRAS.358..363C} to estimate the fractional contribution of the different single stellar populations (SSP) with different ages and metallicities, to the stellar continuum in the spectra.
We can then estimate the mean light-weighted or mass-weighted age and metallicity of the combined stellar population, as well as the total mass.
Dust effects, parametrized by A$_V^{\rm star}$\footnote{The superscript $^{\rm star}$ refers to the nature of the A$_V$ measurement. In this case it is derived using fits to the stellar continuum, while $A_V^{\rm gas}$ refers to the extinction derived from measurements of the Balmer decrement (See \S\ref{sec:ig})}, are modeled as a foreground screen with a \cite{1999PASP..111...63F} 
reddening law assuming R$_V$ = 3.1.
We have used a selection of the SSP model bases, which consists of 66 components with 17 different ages (from 1~Myr to 18~Gyr) and four metallicities (0.2, 0.4, 1.0 and 2.5 \Zsun, where \Zsun=0.02) coming from a slightly modified version of the models of \cite{2003MNRAS.344.1000B}\footnote{See \cite{2007ASPC..374..303B} for more information.}, replacing STELIB by the MILES spectral library \citep{2006MNRAS.371..703S}, Padova 1994 evolutionary tracks, \cite{2003PASP..115..763C} initial mass function (IMF) truncated at 0.1 and 100~\Msun, and new calculations of the TP-AGB evolutionary phase for stars of different mass and metallicity by \cite{2007A&A...469..239M} and \cite{2008A&A...482..883M}. 

\subsection{Ionized gas component}\label{sec:ig}

The {\sc STARLIGHT} fits were subtracted from the observed spectra to obtain pure nebular emission line spectra, and to accurately measure the most prominent emission lines (H$\beta$, [OIII]$\lambda$5007, H$\alpha$, and [NII]$\lambda\lambda$6548, 6583) by means of weighted nonlinear least-squares fit with a single Gaussian plus a linear term. 

The observed ratio of H$\alpha$ and \Hbd\ emission lines provides an estimate of the dust attenuation $A_V^{\rm gas}$ along the line of sight through a galaxy. Assuming an intrinsic ratio  $I$(\Had)/$I$(\Hbd)=2.86, valid for case B recombination with $T=10,000$~K and electron density 10$^2$~cm$^{-3}$ \citep{2006agna.book.....O}, and using \cite{1999PASP..111...63F} Milky Way extinction law, we obtained an estimate of \EBV. Adopting \Rv = $A_V^{\rm gas}$ / \EBV = 3.1 we calculated $A_V^{\rm gas}$. The emission lines previously measured were corrected for the dust extinction before calculating any further measurement. 

To identify AGN contamination in the galaxy centers we used the so-called BPT diagnostic diagram \citep{1981PASP...93....5B, 1987ApJS...63..295V}, a map of $O3\equiv\log_{10}\left(\frac{\textrm{\OIIId}}{\textrm{\Hbd}}\right)$, and $N2\equiv\log_{10}\left(\frac{\textrm{\NIId}}{\textrm{\Had}}\right)$, on which gas ionized by different sources occupies different areas. Two criteria commonly used to separate star-forming (SF) from AGN-dominated galaxies are the expressions in \cite{2001ApJ...556..121K} and \citep{2003MNRAS.346.1055K}. 
However, it should be noted that the latter is an empirical expression,  and {\it bona fide} {\sc H~II} regions can be found in the composite area defined in between the two lines \citep{2014A&A...563A..49S}.
Central spaxels whose spectra had emission falling in the AGN-dominated region according to the criterion of \cite{2001ApJ...556..121K} were excluded, and 2D maps of extinction-corrected H$\alpha$ intensity were created. 

\subsection{{\sc Hii} region segregation}\label{sec:hii}

\begin{figure*}
\centering
\includegraphics[trim=0.2cm 0.2cm 0.2cm 0.2cm, clip=true,width=0.325\textwidth]{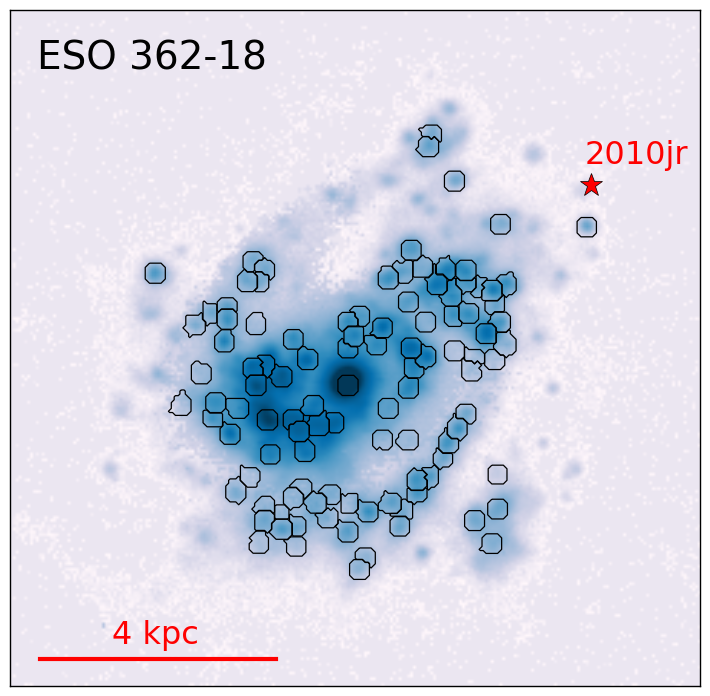}
\includegraphics[trim=0.2cm -1.5cm 0.2cm 0cm, clip=true,width=0.325\textwidth]{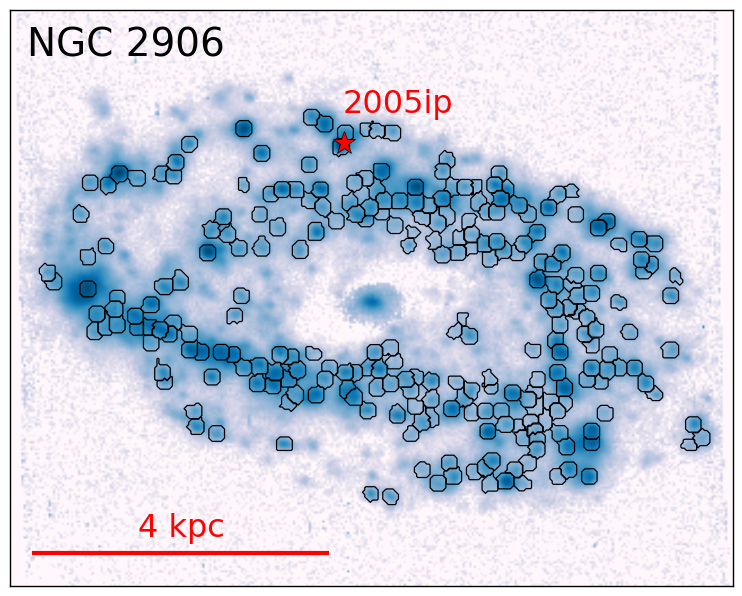}
\includegraphics[trim=0.2cm -4cm 0.2cm 0cm, clip=true,width=0.325\textwidth]{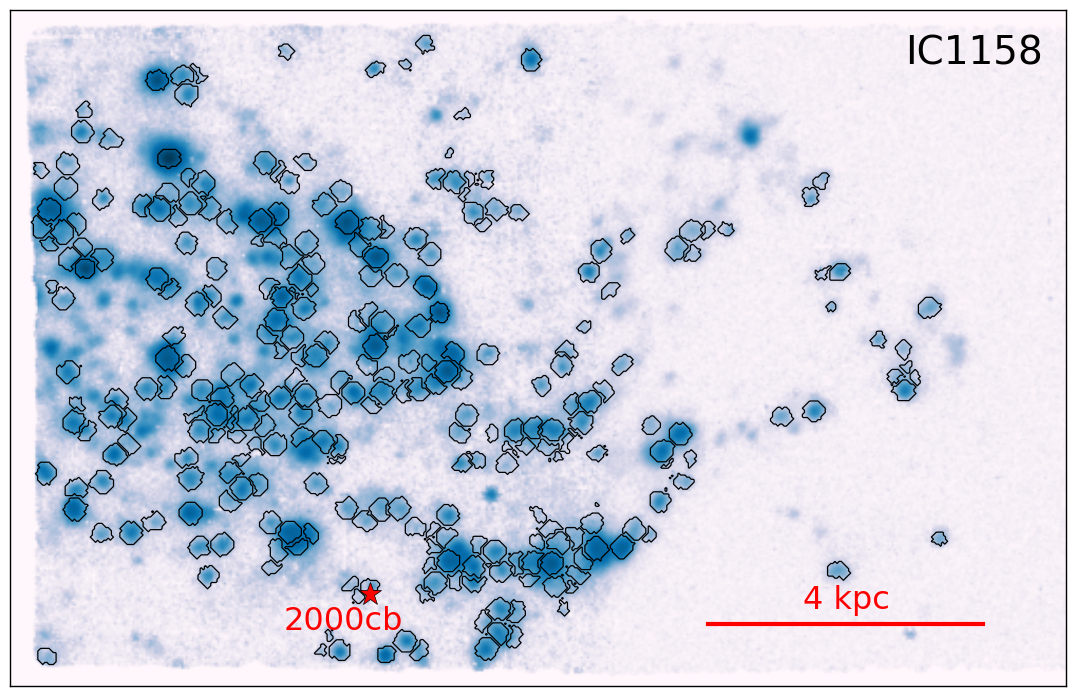}
\includegraphics[trim=0.2cm -5cm 0.2cm 0cm, clip=true,width=0.325\textwidth]{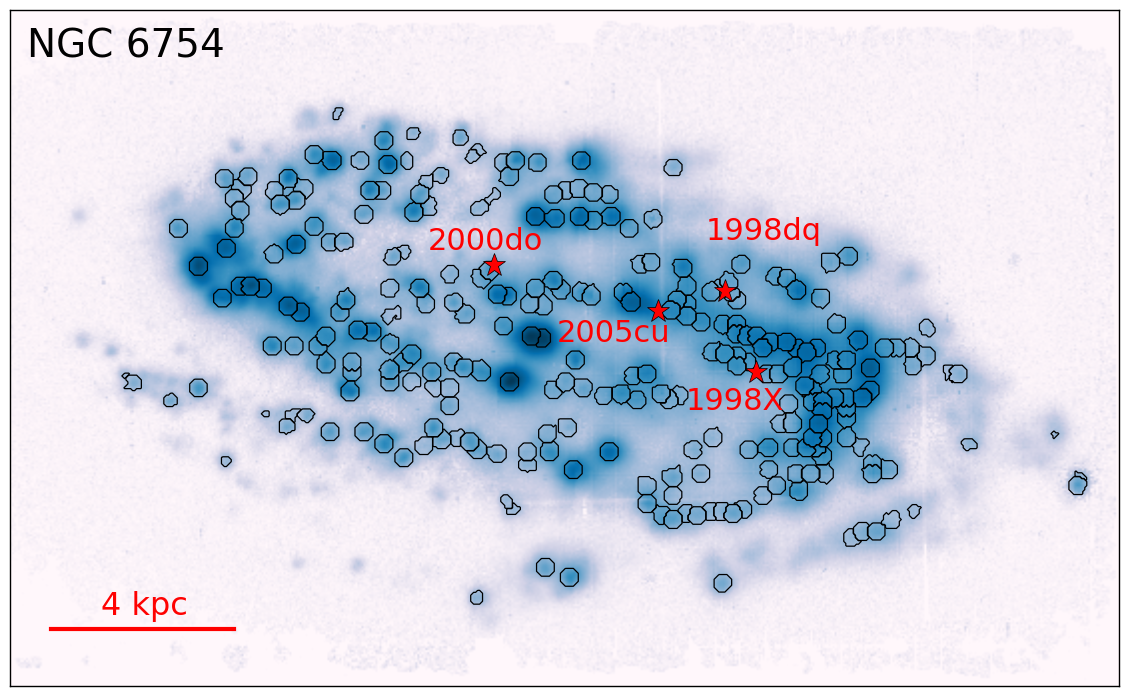}
\includegraphics[trim=0.2cm 0cm 0.2cm 0.2cm, clip=true,width=0.325\textwidth]{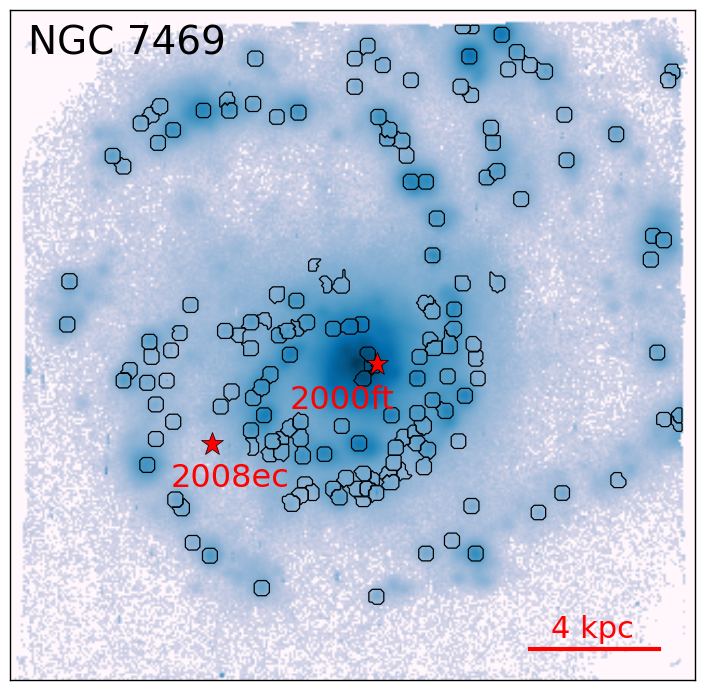}
\includegraphics[trim=0.2cm 0.2cm 0.2cm 0.2cm, clip=true,width=0.325\textwidth]{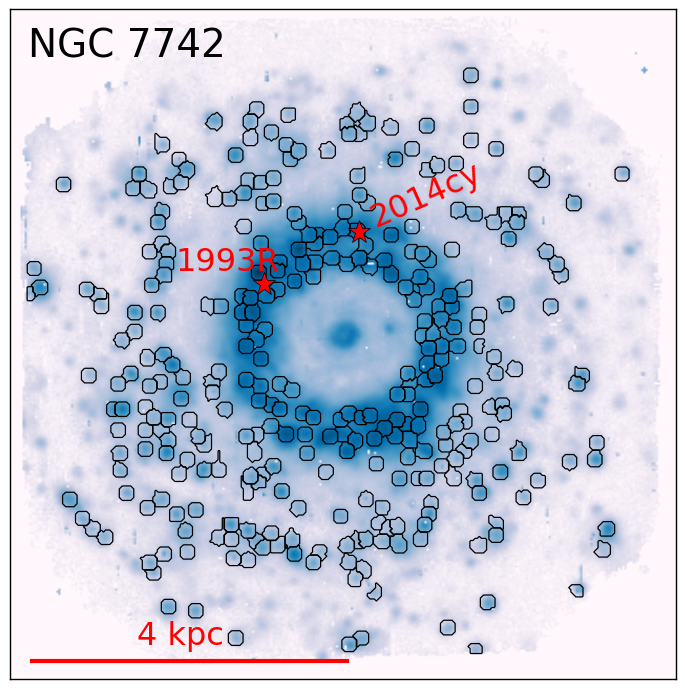}
\caption{Resulting {\sc Hii} regions using {\sc HIIexplorer} and after applying the {\it cleaning} procedure described in \S\ref{sec:hii}.}
\label{fig:hiiexp}
\end{figure*}

The aim of this study is to present a method to characterize the SN parent nebular cluster in comparison to all other populations in the galaxy. 
Our approach here is to select the nebular clump that is closer to the SN explosion, assume that this is the cluster whose properties may be related to the observed properties of the SN and, for CCSN, where the SN progenitor was born.
Then, we compare such properties to the same parameters measured in all other nebular clusters in the galaxy.

We used our extinction-corrected H$\alpha$ maps to select the star-forming {\sc Hii} regions across the galaxy with {\sc HIIexplorer}\footnote{http://www.caha.es/sanchez/HII\_explorer/} \citep{2012A&A...546A...2S}. 
This package detects clumps of higher intensity in the map by aggregating the adjacent pixels until one of the following criteria is reached:  (i) a minimum flux threshold, (ii) a relative flux with respect the peak, or (iii) a radial distance limit.
We put a limit of the median H$\alpha$ emission, a 10\% of the peak flux, and 500 pc as the limits for the three criteria, respectively.
The distance limit takes into account the typical size of {\sc Hii} regions of a few hundreds of parsecs (e.g., \citealt{1997ApJS..108..199G, 2011ApJ...731...91L}). 
The code starts with the brightest pixel, and iterates until no peak with a flux exceeding the median H$\alpha$ emission flux of the galaxy is left. 
This procedure has already been used in \cite{2014A&A...572A..38G} with CALIFA data. In that case, the code did not always select individual {\sc Hii} regions since the physical scale of a real {\sc Hii} region could be significantly smaller than the CALIFA pixel size (1 arcsec$^2$). \cite{2014A&A...561A.129M} estimated that usually from 1 to 6 {\sc Hii} regions were selected in data with that resolution. For MUSE data the upper limit must be lower than in the CALIFA case, since the spatial resolution is, at most, 5 times better (seeing dependent).
Another caveat of the procedure is that it tends to select regions with similar sizes, although real {\sc Hii} regions have different sizes. 
Furthermore, progenitors may often be formed in distinct regions compared to those within which they explode. A full analysis of these issues is beyond the scope of this work, but we note that further work in improving {\sc Hii} region identification through {\sc HIIexplorer} (and other codes) is progressing. Also, in later sections we make comparisons between extracted spectral parameters from {\sc Hii} regions and exact explosion sites, which provide estimations of the uncertainties involved in our technique.

Once the {\sc Hii} regions were identified, the same analysis described in \S\ref{sec:sf} and \S\ref{sec:ig} was performed to the extracted spectra, and a {\it cleaning} procedure was applied to discard any of the segregated regions that resulted from a failure of the code (usually a selection of a region of diffuse gas).
We first, required that the H$\alpha$ EW was higher than 6~\AA to be sure that the emission has a star-formation origin, and then we required a S/N of the H$\beta$ emission line of 3 and that this emission has to be higher than a certain threshold.
Due to the nature of the observations, since they were observed with different characteristics, exposure times, etc., this threshold is different for each galaxy and ranges from 10$^{-18}$ erg sec$^{-1}$ cm$^{-2}$ \AA$^{-1}$ for NGC 7742 that shows the greatest S/N, to the 10$^{-16}$ erg sec$^{-1}$ cm$^{-2}$ \AA$^{-1}$ for IC 1158 which shows the lowest S/N ratio.

After these cuts, the number of {\sc Hii} regions that we are left with is:
IC 1158 (292),
NGC 6754 (423),
ESO 362-18 (91),
NGC 7742 (323),
NGC 7469 (169),
NGC 2906 (241).
Which makes a total of 1539 for 6 galaxies ($\sim 250$ per galaxy on average).
In Figure \ref{fig:hiiexp} we show the resulting {\sc Hii} regions overploted on the H$\alpha$ 2D maps. 
It can be seen that this procedure misses some regions with high H$\alpha$ intensity, but they did not pass the criteria described above. 

We note that we detected contamination from the SN 2005ip remnant in its underlying {\sc Hii} region. This has been removed in the following analysis by fitting broad emissions centered at H Balmer lines. 

\begin{figure*}
\centering
\includegraphics[trim=1cm 0cm 0cm 0cm, clip=true,width=0.9\textwidth]{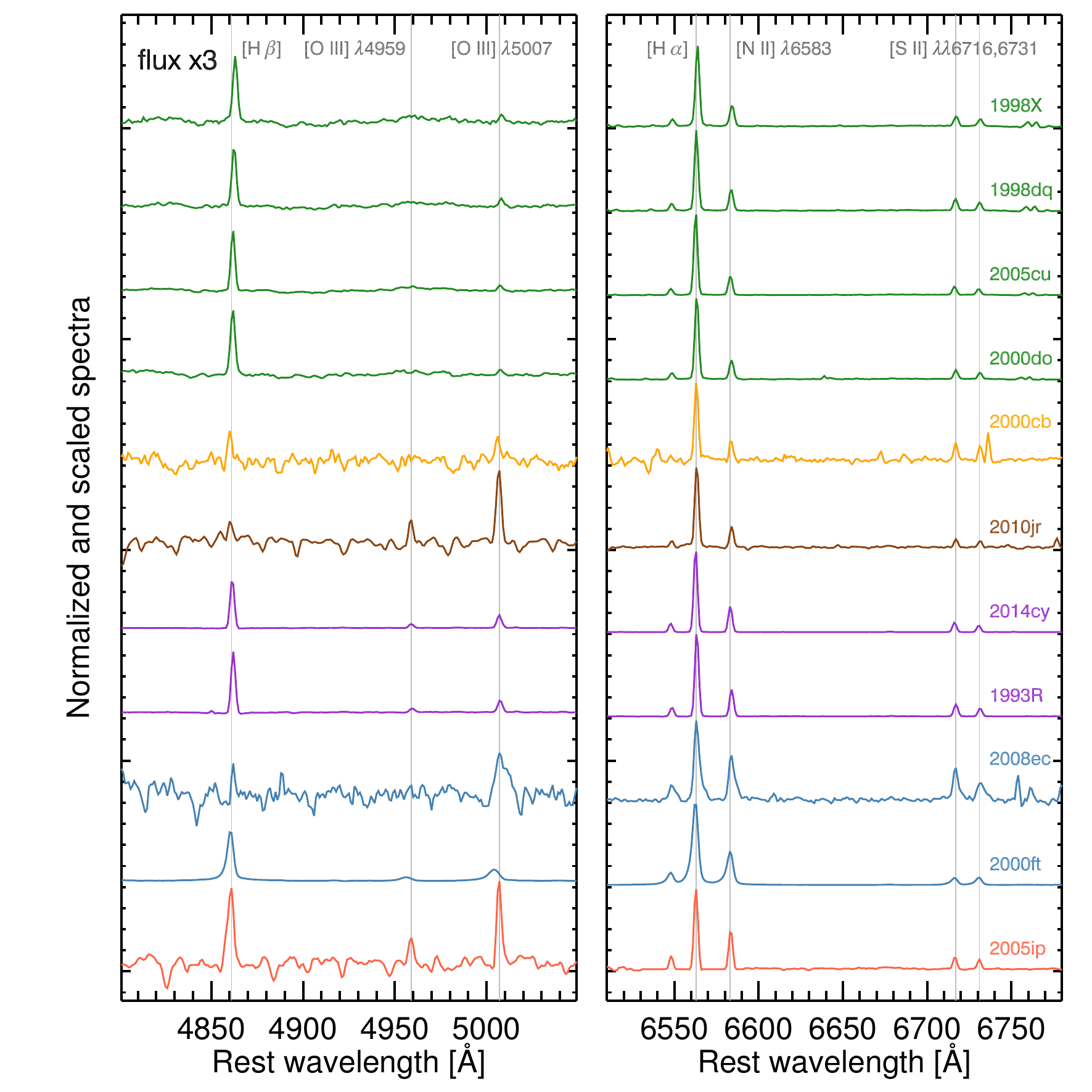}
\caption{Stellar continuum normalized spectra of the nearest {\sc Hii} regions for the 11 SNe. Flux has been normalized to a common H$\alpha$ emission. In the left panel, the normalized flux has been scaled by a factor 3 in order to show clearly the H$\beta$ and the OIII 5$\lambda$5007 lines. The different strength of the emission lines, and the signal-to-noise ratio of the continuum for all spectra are clearly seen.}
\label{fig:hii}
\end{figure*}

\subsection{Studied parameters}

\subsubsection{Star formation rate}

We estimated the ongoing star formation rate (SFR) in all {\sc Hii} regions from the extinction-corrected H$\alpha$ flux $F(\mathrm{H}\alpha)$ using the expression given by \citet{1998ApJ...498..541K}, 
\begin{equation}
{\rm SFR}\,[M_{\sun}\,{\rm yr}^{-1}]=7.9\times 10^{-42}\,L({\rm H}\alpha),
\end{equation}
where $L(\mathrm{H}\alpha)$ is the H$\alpha$ luminosity in units of erg~s$^{-1}$.
\cite{2015arXiv150703801C} demonstrated that the H$\alpha$ luminosity alone can be used as a tracer of the current SFR, even without including UV and IR measurements, once the underlying stellar absorption and the dust attenuation effects have been accounted for.
This measurement was also used to estimate the specific SFR (sSFR=SFR/mass) and the SFR density ($\Sigma$SFR=SFR/area).

\subsubsection{H$\alpha$ equivalent width}

While the H$\alpha$ line luminosity is an indicator of the ongoing SFR, the H$\alpha$ equivalent width (EW) measures how strong the line is compared with the continuum. The continuum light is dominated by old stars, which also contain most of the galaxy stellar mass. 
H$\alpha$ EW can be thought of as an indicator of the strength of the ongoing SFR compared with the past SFR.
In our galaxies, the H$\alpha$ EW was measured from the normalized spectra, resulting from the division of the observed spectra by the {\sc STARLIGHT} fit. It has been shown to be a good proxy for the age of the youngest stellar components \citep{2013AJ....146...30K}.

\subsubsection{Oxygen abundance}\label{sec:oh}

Since oxygen is the most abundant metal in the gas phase and exhibits very strong nebular lines in optical wavelengths, it is usually chosen as a metallicity indicator in ISM studies.
The most accurate method to measure ISM abundances (the so-called direct method) involves determining the ionized gas electron temperature, T$_e$, which is usually estimated from nebular to auroral line intensity ratios, e.g. \OIIIud/\OIIIt~(\citealt{2006A&A...448..955I, 2006A&A...454L.127S}). 
However, in our data the direct method cannot be used since the bluer wavelengths available are around 4700 \AA, and we used instead other strong emission line methods to determine the gas oxygen abundance.
Many such methods have been developed throughout the years.
The theoretical methods are calibrated by matching the observed line fluxes with those predicted by theoretical photoionization models.
The empirical methods, on the other hand, are calibrated against H{\sc II} regions and galaxies whose metallicities have been previously determined by the direct method.
Unfortunately, there are large systematic differences between methods, 
which translate into a considerable uncertainty in the absolute metallicity scale (see \citealt{2010A&A...517A..85L} for a review), while relative metallicities generally agree.
The cause of these discrepancies is still not well-understood, although the empirical methods may underestimate the metallicity by a few tenths of dex, while the theoretical methods overestimate it \citep{2007RMxAC..29...72P, 2010ApJS..190..233M}. 

We used an empirical method based on the O3N2 ratio
\begin{equation}
{\rm O3N2} = \log \left[ \left( {\rm [O III]}~\lambda 5007 / {\rm H}\beta \right) / \left( {\rm [N II]}~\lambda 6583 / {\rm H}\alpha \right) \right],
\end{equation}
to compute the elemental abundances in all galaxies and discussed the results in a relative sense. 
O3N2 was first introduced by \cite{1979A&A....78..200A}, as the difference between the two line ratios used in the BPT diagram (\OIIIHb~and \NIIHa). 
We use here the calibration introduced by \citealt{2013A&A...559A.114M} (hereafter M13),
\begin{equation}
12 + \log~({\rm O/H}) = 8.533 - 0.214 \times{\rm O3N2}. 
\end{equation}
This method has the advantage (over other methods) of being insensitive to extinction due to the small separation in wavelength of the emission lines used for the ratio diagnostics (although this is corrected in our case) and not suffer from differential atmospheric refraction (DAR). 
The uncertainties in the measured metallicities have been computed by explicitly including the statistical uncertainties of the line flux measurements and those in the derived SN host-galaxy reddening, and by properly propagating them into the metallicity determination.


\section{{\sc Hii} region statistics}\label{sec:res}

\begin{table*}\scriptsize
\caption{Average values of the parameters studied in this work for all six galaxies.}           
\label{tab:galres}    
\centering        
\begin{tabular}{lccccc}
\hline\hline   
Galaxy name& H$\alpha$EW                    &log(SFR)                      &log($\Sigma$SFR)                 &log(sSFR)                   & 12+log(O/H)$_{\rm M13,O3N2}$\\
           &[\AA]                           &[M$_{\sun}$ yr$^{-1}$]        &[M$_{\sun}$ yr$^{-1}$ kpc$^{-2}$]&[yr$^{-1}$]                 & [dex]                   \\
\hline                                                                                                                                    
ESO 362-18 & 27.06 {\tiny (-12.76,+15.99) } & -2.26 {\tiny (-0.44,+0.31)}  & -1.29 {\tiny (-0.40,+0.28)}     & -10.28 {\tiny (-0.38,+1.83)}  & 8.43 {\tiny (-0.09,+0.04)} \\
NGC 2906   & 35.38 {\tiny (-16.33,+38.23) } & -3.12 {\tiny (-0.21,+0.40)}  & -1.67 {\tiny (-0.20,+0.37)}     &  -9.40 {\tiny (-1.09,+2.18)}  & 8.58 {\tiny (-0.03,+0.03)}  \\
IC 1158    & 30.51 {\tiny (-12.24,+35.88) } & -3.60 {\tiny (-0.41,+0.34)}  & -2.37 {\tiny (-0.27,+0.28)}     &  -9.98 {\tiny (-0.35,+0.55)}  & 8.41 {\tiny (-0.09,+0.09)}  \\
NGC 6754   & 26.00 {\tiny (-12.67,+26.01) } & -3.29 {\tiny (-0.30,+0.40)}  & -2.35 {\tiny (-0.20,+0.39)}     &  -9.03 {\tiny (-0.71,+0.54)}  & 8.58 {\tiny (-0.04,+0.03)} \\
NGC 7469   & 49.52 {\tiny (-36.27,+107.06)} & -2.52 {\tiny (-0.44,+0.71)}  & -1.80 {\tiny (-0.38,+0.67)}     &  -9.30 {\tiny (-0.42,+0.46)}  & 8.50 {\tiny (-0.02,+0.03)} \\
NGC 7742   & 39.80 {\tiny (-22.30,+48.90) } & -3.32 {\tiny (-0.29,+0.83)}  & -1.67 {\tiny (-0.28,+0.87)}     &  -8.16 {\tiny (-0.38,+0.53)}  & 8.52 {\tiny (-0.04,+0.05)}  \\
 \hline
\end{tabular}
\end{table*}

The nearest SN {\sc Hii} region was selected by measuring the deprojected distance from the SN position to the center of all {\sc Hii} regions, and looking for the lowest value. For the deprojection of distances we used the H$\alpha$ velocity map and analyzed the kinematics.
SN positions have been determined from galaxy coordinates applying the reported offsets in discovery telegrams. 
For SNe II, and given the short elapse (few tens of Myr) between the birth of the progenitor star and the supernova explosion, these were considered to be the parent {\sc Hii} regions.
In the case of SNe Ia, their long delay times (100 Myrs to Gyrs) mean that associating any given environment directly to the SN progenitor population is much more problematic given that progenitors could have moved a considerable distance before explosion. However, it is still possible that interesting results can be obtained in analysing these explosion sites (see e.g. \citealt{2013A&A...560A..66R,2015MNRAS.448..732A,2015Sci...347.1459K}), and hence we proceed in this fashion.
In Figure \ref{fig:hii} we present the spectra of the 11 selected nearest SN {\sc Hii} regions.

IFS data allows a proper characterization of the SN environment with respect to other regions of the galaxy. 
In the following we construct galaxy-wide distributions of the H$\alpha$ EW, SFR, $\Sigma$SFR, sSFR, and Oxygen abundance, measured in all {\sc Hii} regions of the galaxy, and position the nearest SN {\sc Hii} region values in those distributions.
In Table \ref{tab:galres} we give the average values of the parameters for all galaxies, while Table \ref{tab:snres} lists the nearest SN {\sc Hii} region values.
Table \ref{tab:avg} contains the results split in two groups: we give for SNe II and SNe Ia the average value of each parameter presented in this work and the average position in the cumulative distributions.
Figures \ref{fig:hiiew}, \ref{fig:hiisfr}, and \ref{fig:hiioh} show these distributions. Overploted are the values at the nearest SN {\sc Hii} regions, and the averages of these SNe II and SNe Ia subsamples, which will be discussed below.

\subsection{H$\alpha$ equivalent width}

\begin{figure*}
\centering
\includegraphics[trim=0.7cm 0.1cm 0.4cm 0.4cm, clip=true,width=\textwidth]{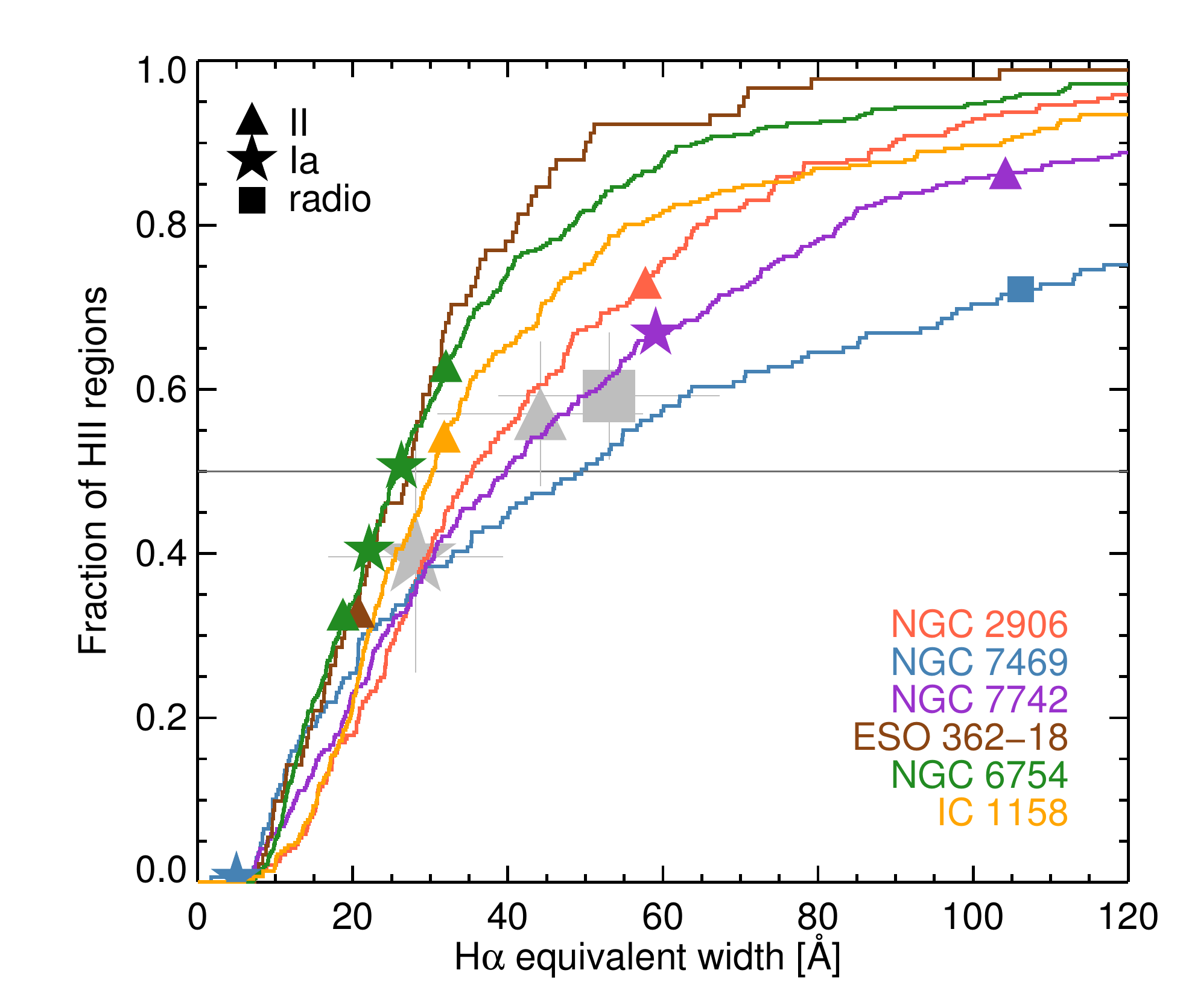}
\caption{Cumulative distribution of the H$\alpha$ equivalent width of all {\sc Hii} regions in the galaxies. Each colored line represents one of the six galaxies, while symbols (stars for SN Ia, triangles for SNII, and squares for the radio SN) determines the position and the H$\alpha$EW of the nearest SN {\sc Hii} region. A grey horizontal line crosses the distributions at their median values. Three big grey symbols in the background, represent the average positions of the SNe Ia (big star) and SNe II (big square for the 7 SNe II, and a big triangle for the 6 SNe II without the radio SN) and their errors represent the standard error of the mean.}
\label{fig:hiiew}
\end{figure*}

All cumulative distributions (CD) in Figure \ref{fig:hiiew} increase steadily up to H$\alpha$EW values around $\sim$30 \AA, to then keep increasing shallowly up to $\sim$250 \AA. This correspond to distributions peaking at H$\alpha$EW $\sim$30 \AA~and having a wing that goes to higher values.
The mean H$\alpha$EW values for the six galaxies are quite similar, from 26.00 ($^{-12.76}_{+26.01}$) \AA~for NGC 6754 to 49.52 ($^{-36.27}_{+107.06}$) \AA~for NGC 7469.
Since, under certain assumptions, this can be used as proxy for the age of the stellar populations present in the galaxies (see \S\ref{sec:hanin}), we can order our galaxies from the one having, on average, older {\sc Hii} regions to the younger. 

Strong conclusions cannot be taken from the results of nearest SN {\sc Hii} regions, given the low numbers on our sample.
Nevertheless, they are above the median H$\alpha$EW of their host galaxies (represented by a horizontal line in Figure \ref{fig:hiiew}), except for four cases: SN 2008ec in NGC 7469, SN 2010jr in ESO 362-18, and SN 2000do and SN 1998X in NGC 6754.
Two of them are type Ia SNe and the other two are type II SNe.
The lowest value is at SN 2008ec {\sc Hii} region (6.99 $\pm$ 0.32 \AA), and the highest value is at the radio SN 2000ft  {\sc Hii} region (106.16 $\pm$ 3.22 \AA). Excluding this radio SN, 2014cy {\sc Hii} region would be the one showing the highest value of 104.17 (0.84) \AA.

\begin{figure}
\centering
\includegraphics[trim=1.3cm 0.3cm 0.6cm 0.1cm, clip=true,width=0.9\columnwidth]{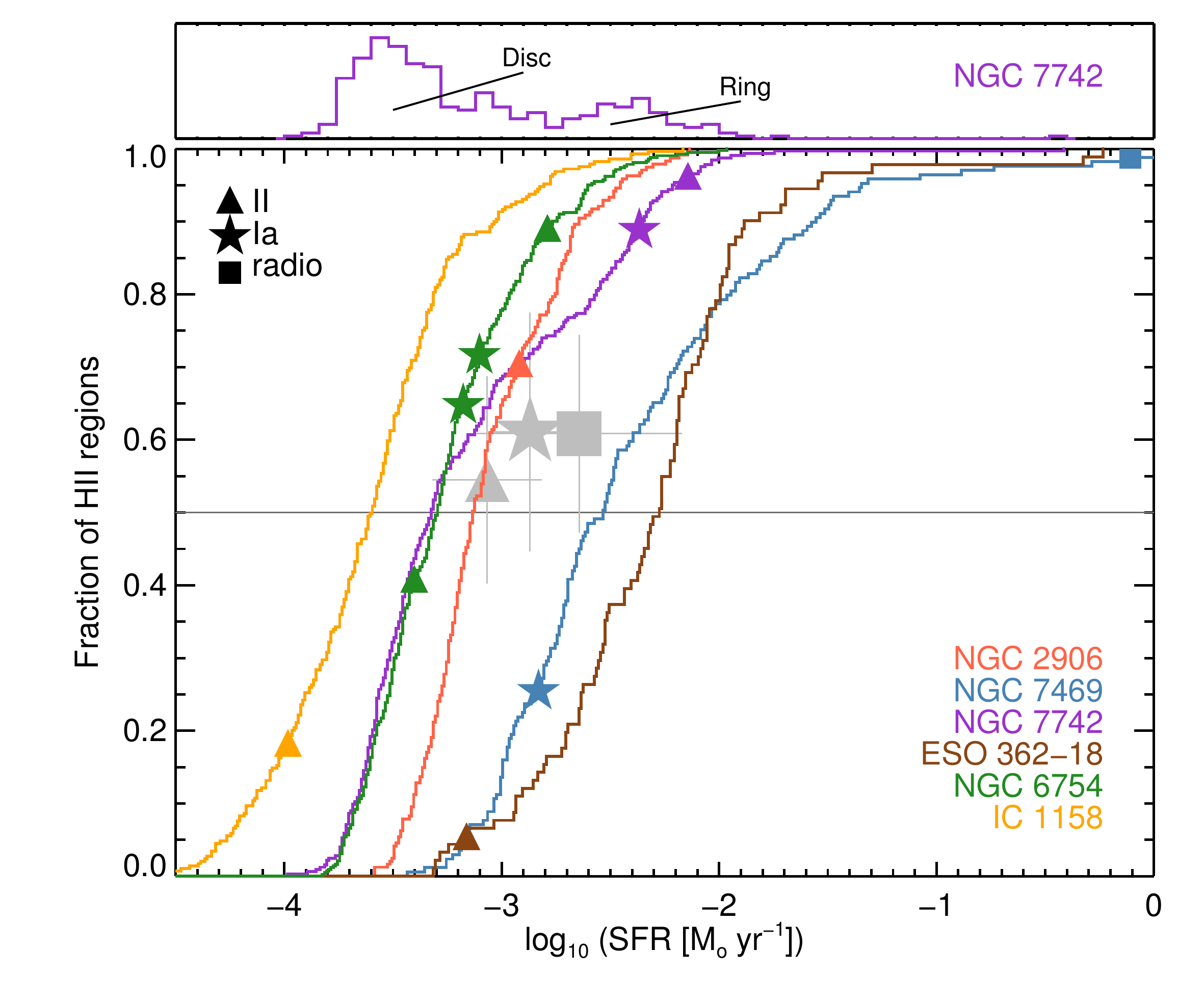}\\
\includegraphics[trim=1.3cm 0.3cm 0.6cm 0.1cm, clip=true,width=0.9\columnwidth]{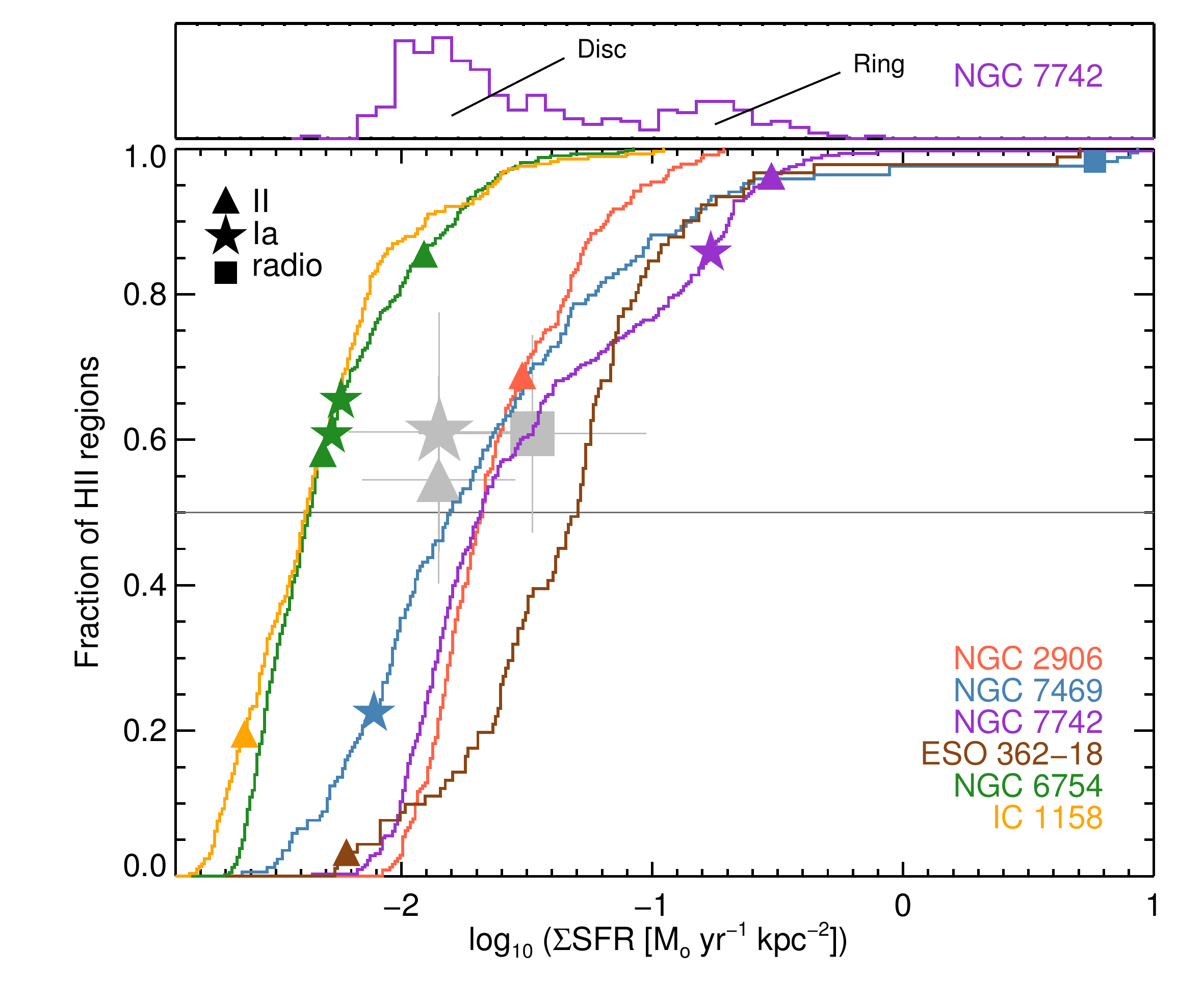}\\
\includegraphics[trim=1.3cm 0.3cm 0.6cm 0.6cm, clip=true,width=0.9\columnwidth]{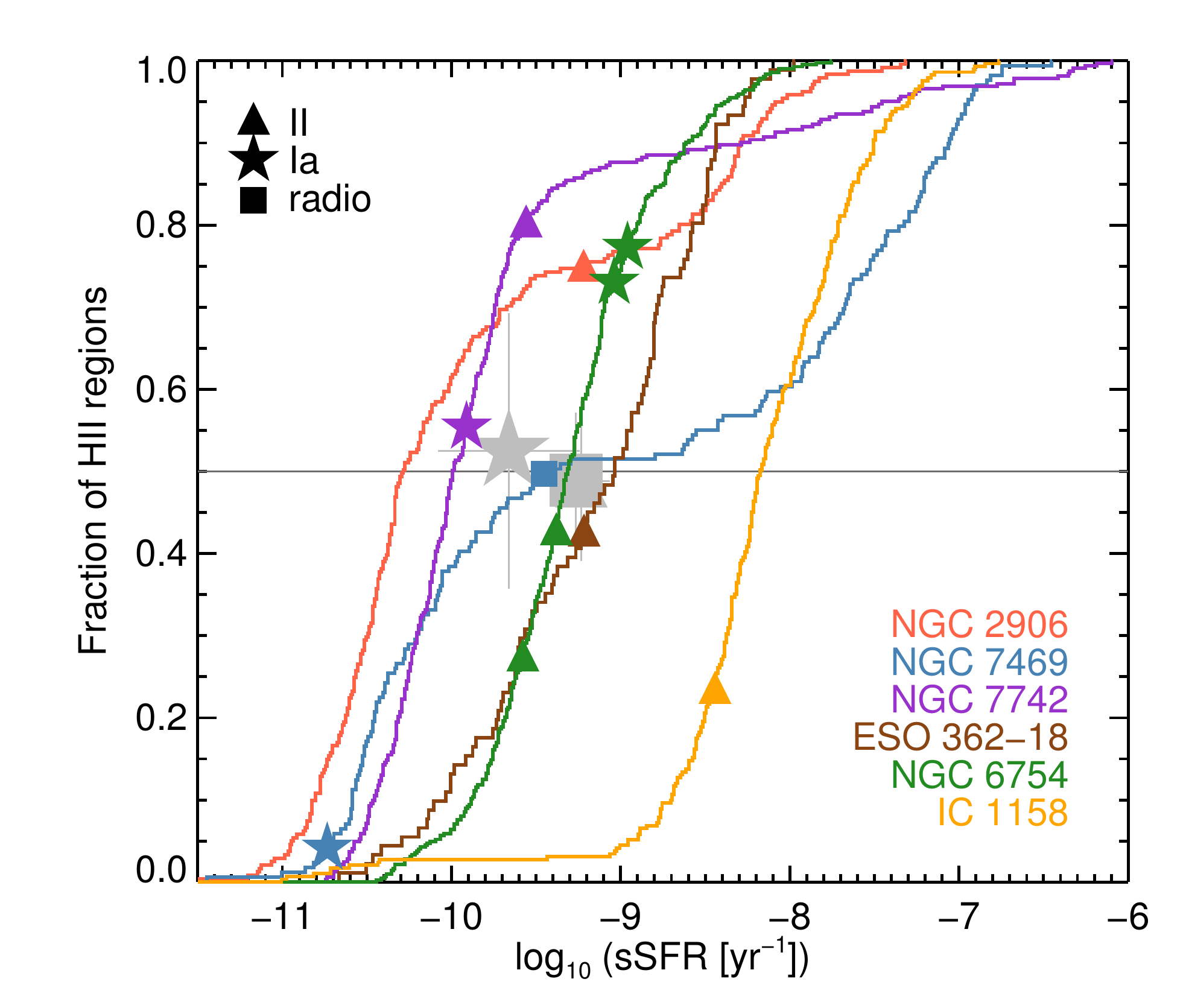}
\caption{Cumulative distribution of the SFR, SFR density, and specific SFR of all {\sc Hii} regions in the galaxies. Each colored line represents one of the six galaxies, while symbols (stars for SN Ia, triangles for SNII, and squares for the radio SN) shows the position and the measurement of the nearest SN {\sc Hii} region. A grey horizontal line crosses the distributions at their median values. Three big grey symbols in the background, represent the average positions of the SNe Ia (big star) and SNe II (big square for the 7 SNe II, and a big triangle for the 6 SNe II without the radio SN) and their errors represent the standard error of the mean.}
\label{fig:hiisfr}
\end{figure}

\begin{figure}
\centering
\includegraphics[trim=0cm 0.2cm 0cm 0.3cm, clip=true,width=\columnwidth]{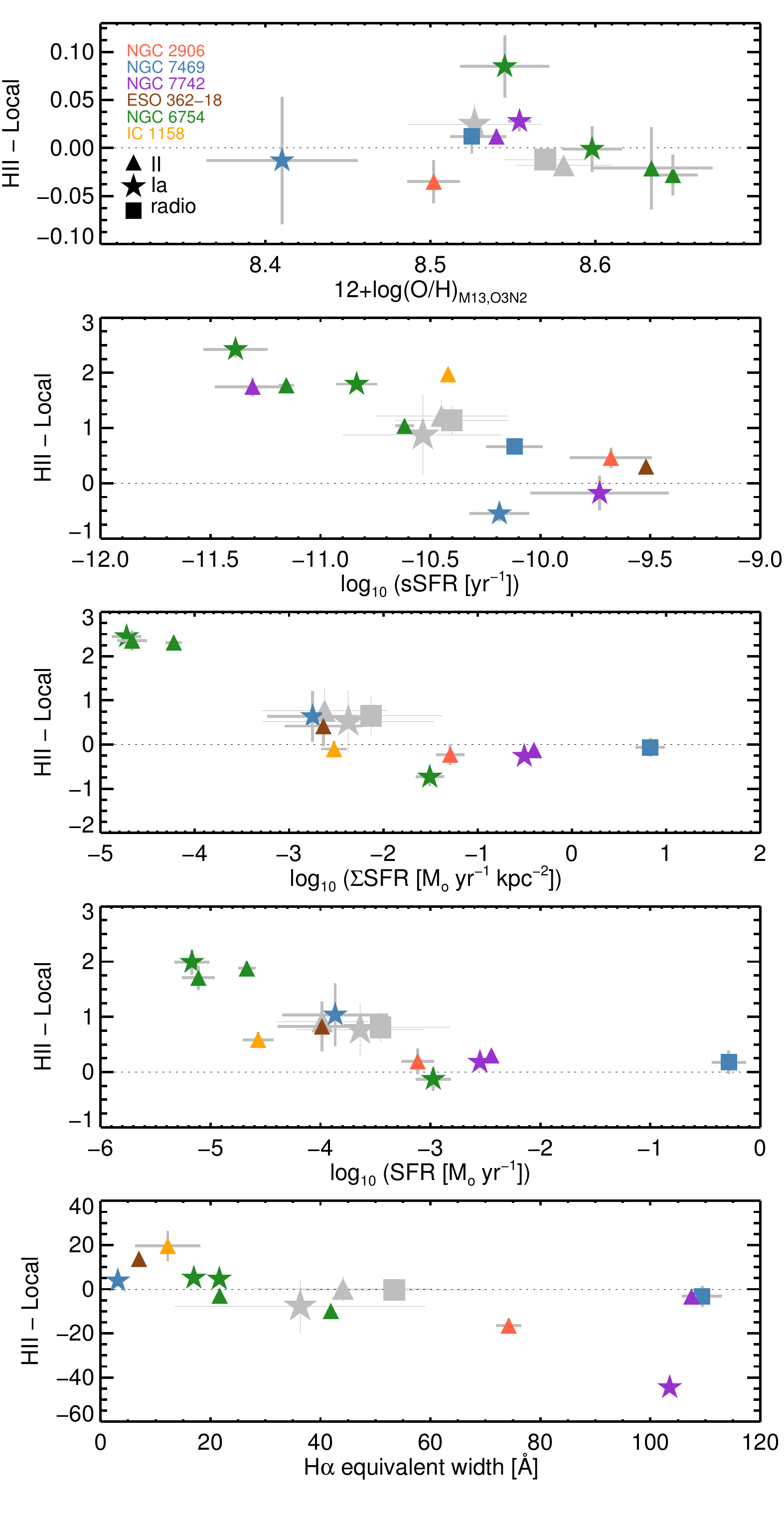}
\caption{Difference between the measurement of the parameters at the nearest {\sc Hii} region and at the SN position, with respect the value at the SN position.}
\label{fig:sc}
\end{figure}

The average H$\alpha$ EW for SNe II in our sample is 53.05 (14.25) \AA, while for SNe Ia is lower 28.10 (11.30) \AA. On average, the parent SNe II {\sc Hii} regions have H$\alpha$ EW values above the median of all {\sc Hii} regions in their galaxies ($\avg{\rm CD}$=0.59 $\pm$ 0.08), while the nearest SNe Ia {\sc Hii} regions have lower values ($\avg{\rm CD}$=0.40 $\pm$ 0.14). 
Given the low number statistics involved in the current study, any significant conclusions are premature. However, the main aim of this paper is to show how these wide-field IFS data can be used to analyse SNe environments in new statistical ways. Plots such as those presented in Figures \ref{fig:hiiew} to \ref{fig:sc} show the potential of galaxy-wide environment studies to differentiate SN progenitor properties by asking where within the distribution of host galaxy stellar population properties SNe are found to explode.

\subsubsection{Age inferred from H$\alpha$ EW}\label{sec:hanin}

The H$\alpha$EW provides an estimate of the age of a stellar population. In a stellar population born within a single, instantaneous starburst, the EWs of the Balmer lines decrease with increasing stellar population age. This is due to the constantly decreasing number of massive ionizing stars, which are responsible for the nebular emission. As a result, the line emission becomes less prominent compared to the stellar continuum, thus EW decreases. Comparing the observed H$\alpha$EW of the {\sc Hii} regions of the SNe with Starburst99 SSP models \citep{1999ApJS..123....3L}, we estimated the age of the associated stellar populations.
It should be noted that the measured H$\alpha$EW is dependent on the strength of the underlying continuum, which contribution could come from older populations not related to the HII region. For this reason we consider our measurements as lower limits for EWs and upper limits for stellar ages.
Assuming single starburst and adopting the respective metallicity values, this method gives an age of 6.13 (0.02) Myr for SN 2000ft and 13.45 (0.24) Myr for SN 2008ec. The remaining SNe would have ages in between these two SNe. The average SNe II parent cluster age, 7.23 (0.48) Myr, is lower than the corresponding age of SNe Ia clusters, 8.72 (1.62) Myr. This would indicate that SNe II are more associated with more massive stars compared to SNe Ia. 

\subsection{Star Formation Rate}

We show in Figure \ref{fig:hiisfr} the star formation rate (top), the star formation rate density (middle), and the specific star formation rate (bottom) distributions for all the {\sc Hii} regions in our six galaxies.
The average SFR values range from -3.60 ($^{+0.41}_{-0.34}$) dex \Msun~yr$^{-1}$ for IC 1158 to -2.26 ($^{+0.44}_{-0.31}$) dex \Msun~yr$^{-1}$ for ESO 362-18, while for the $\Sigma$SFR goes from -2.37 ($^{+0.27}_{-0.28}$) dex \Msun~yr$^{-1}$ kpc$^{-2}$ for IC 1158 to -1.29 ($^{+0.40}_{-0.28}$) dex \Msun~yr$^{-1}$ kpc$^{-2}$ for ESO 362-18. The lowest sSFR is -10.28 ($^{+0.27}_{-0.28}$) yr$^{-1}$ for ESO 362-18, and the highest is -8.16 ($^{+0.38}_{-0.53}$) yr$^{-1}$ for NGC 7742.
Although NGC 6754 has a similar average SFR and sSFR than NGC 2906 and NGC 7742, we see that the two latter have significantly higher $\Sigma$SFR due to their smaller physical size of the {\sc Hii} region. This stresses the importance of measuring the SFR density to properly compare galaxies of different sizes and distances, which correspond to spaxels of different size.
IC 1158 is the galaxy showing the lowest values in all indicators but in sSFR.
Since the sSFR is the ongoing SFR per unit galaxy stellar mass, it gives an indication on the contribution of the current star formation to the galaxy growth.
This indicates that a significant fraction (with respect to the other 5 galaxies) of its mass has been recenlty created.
Moreover, we note that this galaxy, and NGC 6754, also required two pointings to cover part of it. On the other hand, the four galaxies that fitted in one pointing have $\Sigma$SFR higher than -2 dex \Msun~yr$^{-1}$ kpc$^{-2}$.

SFR and $\Sigma$SFR CDs presented in Figure \ref{fig:hiisfr} for NGC 7742 have shapes that correspond to a typical two peak distribution. They are represented in the upper panel of both CDs. The separation of the two peaks is around $\sim$-2.8 dex \Msun~yr$^{-1}$ in the SFR distribution, and at $\sim$-1.3 dex \Msun~yr$^{-1}$  kpc$^{-2}$ in the $\Sigma$SFR distribution, where the bump in the CD that corresponds to the second peak is around $\sim$-2.5 dex \Msun~yr$^{-1}$ in the SFR distribution, and at $\sim$-1.0 dex \Msun~yr$^{-1}$ kpc$^{-2}$ in the $\Sigma$SFR distribution.
The first peak, with lower SFR and $\Sigma$SFR would correspond to the {\sc Hii} regions across the disc of the galaxy, while the second and shortest peak centered at higher SFR and $\Sigma$SFR values would represent those {\sc Hii} regions located on the circular ring where the two nearest SN {\sc Hii} regions are located (See Figure \ref{fig:hiiexp}).
This is underlying that the star formation is more important on the ring when compared to the disc.
All other galaxies follow shapes of skewed one-peak distributions.

Interestingly, and with different strengths, all sSFR CDs show the same behavior of a two-peak distribution. In NGC 2906, NGC 7469 and NGC 7742 this is clearly seen. The CDs grow up to a level where they become flatter (this corresponds to the first peak with {\sc Hii} region low sSFR values) to then rise again (this corresponds to the second peak with higher sSFR). While in {\sc Hii} regions that contribute to the second peak a significant contribution to the mass is being created in the last few Myr (the typical age of the {\sc Hii} region), in regions contributing to the first peak the old stars are more important. 
Two out of four SN Ia occur in the first distribution, while all SN II are located either on the second distribution or in between the two (close to the flatter part of the CDs).
This parameter also indicates that SNe II are more associated with regions where the current star formation is more important with respect to all the populations present in the environment.

Again the radio SN 2000ft parent {\sc Hii} region presents the highest SFR and $\Sigma$SFR parameters, -0.11 (0.02) dex \Msun~yr$^{-1}$ and 0.76 (0.02) dex \Msun~yr$^{-1}$ kpc$^{-2}$.
Discarding this object, the next would be again SN 2014cy  parent {\sc Hii} region with -2.14 (0.01) dex \Msun~yr$^{-1}$ and -0.53 (0.01) dex \Msun~yr$^{-1}$ kpc$^{-2}$.
The one showing the lowest values in both parameters is SN II 2000cb in IC 1158.
On the other hand, this SN is the one with highest sSFR -8.44 (0.05) dex~yr$^{-1}$, while SN Ia 2008ec in NGC7469 has the lowest value -10.74 (0.03)  dex~yr$^{-1}$. 

While 4 SNe are below the median SFR of their galaxy, only 3 of them remain at lower values than the median when considering the $\Sigma$SFR: 
SN 1998X parent {\sc Hii} region is above the average when the $\Sigma$SFR is considered.
For the sSFR we considered more important if the nearest SN {\sc Hii} region was closer to the first or the second peak of the sSFR distribution, than if it was above or below the median sSFR of the galaxy.

\subsection{Oxygen abundance}

\begin{figure*}
\centering
\includegraphics[trim=0.7cm 0.1cm 0.4cm 0.4cm, clip=true,width=\textwidth]{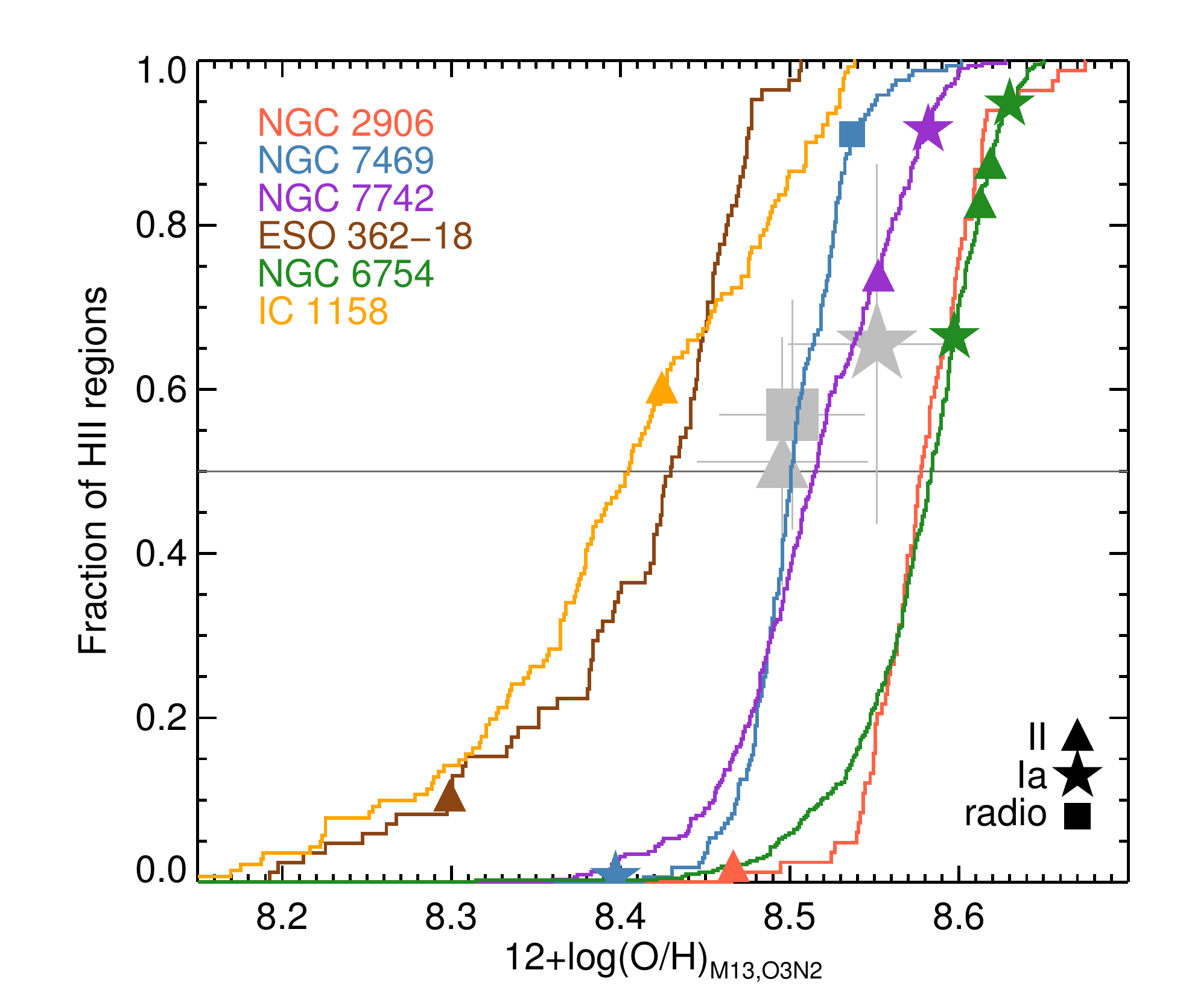}
\caption{Cumulative distribution of the oxygen abundance of all {\sc Hii} regions in the galaxies. Each colored line represents one of the six galaxies, while symbols (stars for SN Ia, triangles for SNII, and squares for the radio SN) determines the position and the oxygen abundance of the nearest SN {\sc Hii} region. A grey horizontal line crosses the distributions at their median values. Three big grey symbols in the background, represent the average positions of the SNe Ia (big star) and SNe II (big square for the 7 SNe II, and a big triangle for the 6 SNe II without the radio SN) and their errors represent the standard error of the mean.}
\label{fig:hiioh}
\end{figure*}

Figure \ref{fig:hiioh} shows the distribution of the oxygen abundances measured with the O3N2 empirical method using the M13 calibration.
Median oxygen abundances of the six galaxies in our sample range from 8.41 (0.09) dex for IC 1158, to 8.58 (0.03) dex for NGC 6754.

Three SNe (2 II and 1 Ia) have lower oxygen abundances than the median value of all the {\sc Hii} regions in their galaxies, and most of them explode close to the metal-higher end {\sc Hii} regions of their galaxies.

SN II 2010jr in ESO 362-18 is the object with the lowest nearest {\sc Hii} oxygen abundance, 8.299 (0.012), while SN Ia 2000do in NGC 6754 is the one with the highest local metallicity 8.630 (0.018).

Average SNe Ia nearest {\sc Hii} region abundances are higher than SNe II by 0.05 dex, and while the latter are located on average at the median abundance value of their host galaxies, SN Ia nearest {\sc Hii} regions are on average at higher metallicities than the median ($\avg{CD}$=0.66 $\pm$ 0.22).
This is in agreement to what Galbany et al. in prep found studying the SN host galaxies observed by the CALIFA survey. While SNe II locations have metallicities similar to the integrated value of their hosts, SNe Ia locations tend to have higher values.
When the radio SN II is included in the average SN II metallicity the difference in metallicity between SN Ia and SN II is not affected, but the average position in the CDs is slightly shifted to higher values (0.57 $\pm$0.14).

\begin{table*}\scriptsize
\caption{Results of the parameters studied in this work at the nearest SN {\sc Hii} region position and at the SN position.}           
\label{tab:snres}    
\centering        
\begin{tabular}{lccccccc}
\hline\hline   
SN name&type        & H$\alpha$EW   & Age            &SFR                   &$\Sigma$SFR                      &sSFR           & 12+log(O/H)$_{\rm M13,O3N2}$\\
       &            &[\AA]          & [Myr]          &[M$_{\sun}$ yr$^{-1}$]&[M$_{\sun}$ yr$^{-1}$ kpc$^{-2}$]&[yr$^{-1}$]    & [dex]                   \\
\hline                                                                                                          
\multicolumn{8}{c}{nearest {\sc Hii} region} \\                                                                
\hline                                                                                                          
2005ip &IIn         &  57.74 (2.26) &  6.96 (0.03)   &1.20(0.05) e-03       & 3.03(0.13) e-02                 &6.04 (0.25) e-10& 8.467 (0.016) \\
2000ft &II radio    & 106.16 (3.22) &  6.13 (0.02)   &7.79(0.26) e-01       & 5.82(0.20) e+00                 &3.52 (0.12) e-10& 8.537 (0.012) \\
2008ec &Ia          &   6.99 (0.32) & 13.45 (0.24)   &1.48(0.11) e-03       & 7.77(0.58) e-03                 &1.84 (0.14) e-11& 8.397 (0.048) \\
1993R  &Ia 91bg-like&  59.06 (0.63) &  6.36 (0.01)   &4.28(0.06) e-03       & 1.71(0.02) e-01                 &1.23 (0.02) e-10& 8.582 (0.008) \\
2014cy &II          & 104.17 (0.84) &  6.13 (0.01)   &7.17(0.08) e-03       & 2.98(0.03) e-01                 &2.76 (0.03) e-10& 8.552 (0.006) \\
2010jr &II          &  20.72 (0.87) &  9.36 (0.47)   &6.89(0.33) e-04       & 6.05(0.29) e-03                 &6.07 (0.29) e-10& 8.299 (0.012) \\
2000do &Ia          &  22.09 (0.55) &  8.16 (0.88)   &7.90(0.26) e-04       & 5.74(0.19) e-03                 &1.10 (0.04) e-09& 8.630 (0.018) \\
2005cu &II          &  32.01 (0.59) &  6.68 (0.01)   &1.62(0.04) e-03       & 1.23(0.03) e-02                 &2.64 (0.07) e-10& 8.619 (0.015) \\
1998dq &Ia          &  26.26 (0.79) &  6.89 (0.11)   &6.65(0.26) e-04       & 5.28(0.20) e-03                 &9.14 (0.36) e-10& 8.597 (0.016) \\
1998X  &II          &  18.73 (0.56) &  8.65 (0.89)   &3.96(0.15) e-04       & 4.87(0.19) e-03                 &4.17 (0.16) e-10& 8.613 (0.022) \\
2000cb &II 87A-like &  31.79 (3.47) &  6.68 (0.09)   &1.04(0.12) e-04       & 2.37(0.77) e-03                 &3.61 (0.41) e-09& 8.424 (0.019) \\
 \hline                                                                                                                       
 \multicolumn{8}{c}{SN location} \\                                                                                          
\hline                                                                                                                        
2005ip &IIn         &  74.23 (2.29) & 6.28 (0.01)    &7.65(0.27) e-04       & 5.13(0.18) e-02                 &2.09 (1.12) e-10 & 8.502 (0.016) \\
2000ft &II radio    & 109.39 (3.67) & 6.11 (0.02)    &5.17(0.19) e-01       & 6.79(0.25) e+00                 &7.62 (2.64) e-11 & 8.525 (0.013) \\
2008ec &Ia          &   3.14 (0.29) &18.36 (2.86)    &1.36(0.16) e-04       & 1.78(0.21) e-03                 &6.51 (2.40) e-11 & 8.410 (0.046) \\
1993R  &Ia 91bg-like& 103.51 (0.99) & 6.14 (0.01)    &2.82(0.01) e-03       & 3.13(0.04) e-01                 &1.86 (1.98) e-10 & 8.554 (0.007) \\
2014cy &II          & 107.46 (0.84) & 6.12 (0.01)    &3.58(0.01) e-03       & 3.96(0.04) e-01                 &4.91 (2.39) e-12 & 8.540 (0.006) \\
2010jr &II          &   6.98 (0.60) &12.39 (0.10)$^a$&1.03(0.10) e-04       & 2.31(0.23) e-03                 &3.02 (0.01) e-10 &   ---         \\
2000do &Ia          &  16.97 (0.52) &10.05 (0.66)    &1.06(0.04) e-03       & 3.11(0.11) e-02                 &4.11 (1.65) e-12 & 8.545 (0.027) \\
2005cu &II          &  41.88 (0.69) & 6.53 (0.01)    &2.14(0.04) e-05       & 5.98(0.12) e-05                 &2.41 (0.25) e-11 & 8.647 (0.015) \\
1998dq &Ia          &  21.60 (0.64) & 8.32 (1.02)    &6.78(0.25) e-06       & 1.89(0.07) e-05                 &1.46 (0.35) e-11 & 8.598 (0.018) \\
1998X  &II          &  21.66 (0.58) & 8.31 (1.01)    &7.76(0.27) e-06       & 2.16(0.08) e-05                 &6.99 (0.58) e-12 & 8.634 (0.037) \\
2000cb &II 87A-like &  12.21 (5.97) &11.16 (2.32)$^b$&2.71(1.02) e-05       & 3.00(1.13) e-03                 &3.80 (0.08) e-11 &   ---         \\
\hline
\end{tabular}

$^a$assuming a metallicity of 8.529 (0.042) measured with the M13 N2 method.
$^b$Result assuming Z$_{\sun}$. It would be 11.12 (2.58) Myr assuming Z$_{\rm LMC}$.
\end{table*}

\subsection{Comparison with local site}

Above, we compared the parameters measured from the nearest SN {\sc Hii} region to all other {\sc Hii} regions in the galaxy. 
For SN II, we assumed that the closest {\sc Hii} region corresponds to the parent nebular cluster where the SN progenitor star was born, and the properties measured might be similar to those of the progenitor.
For SN Ia, on the other hand, the nearest {\sc Hii} region do not have to correspond to the progenitor parent cluster due to the older age of their progenitors and the fact that they have travelled across the galaxy since their formation until the SN. However, the study of the properties of SN Ia environments can give insights on the characteristics of the explosion that might be useful, for example, for cosmological analyses.

All the parameters studied in this work come from the measurement of the ionized gas emission. 
While CC SNe are only found in late-type galaxies close to regions with signs of recent star formation, SN Ia are also found in regions with no signs of star formation and in early-type galaxies. 
Then, differences between the parameters at the SN II locations when compared to the closest {\sc Hii} region, are not expected to be very significant, while depending on the distance to the {\sc Hii} region, the locations of SN Ia can have very different properties.

Here, we extracted aperture spectra at all SN locations with a diameter that equals the seeing of each observation (See Table \ref{tab:obs}), and compared to the measurement at the nearest {\sc Hii} region to see whether the differences introduce a bias in the determination of the progenitor properties.
In Table \ref{tab:snres} we also give the resulting measurements of the same parameters presented in the previous section, for this local SN environment spectra.

Most of our SNe are located inside one of the segregated {\sc Hii} regions, or very close to one of them ($<$1 arcsec), so the differences in the parameters are not expected to be very high. The exceptions are SN II 2010jr in ESO 362-18 and SN Ia 2008ec in NGC 7469, that are further than 2 arcsec.

When looking at the H$\alpha$EW differences, we find that 5 out of 7 SN II tend to have higher H$\alpha$EW values at their positions compared to the respective parent nebular cluster. 
In contrast, SN 2010jr and SN 2000cb have lower values at their positions. The former is one of the two SN located far from a {\sc Hii} region, and the latter was classified as a peculiar event (1987A-like). 
In comparison, the opposite happens for SN Ia, the local value for 3 out of 4 SN tend to be lower than that of the closer {\sc Hii} region, with the exception is SN 1993R that has almost twice H$\alpha$ EW locally. 
This is a peculiar subluminous event that exploded in the ring of NGC 7742. Looking at the H$\alpha$ map in Figure \ref{fig:hiiexp}, one can attribute the reason of this high local value to the fact that a very strong H$\alpha$ emission from a closer {\sc Hii} region is contributing to the aperture spectrum extracted centered at the SN position.

Since this parameter is correlated to the age of the stellar population, this trend roughly points out that the location of the CCSN are younger with respect of their corresponding {\sc Hii} regions, while the locations of SN Ia contain older population than their nearest nebular clusters.

\begin{table}
\caption{Average values of the parameters studied in this workmeasured in the nearest SN {\sc Hii} region, and of the position in the CD of all {\sc Hii} regions in the galaxy, for the two SN types.}     
\label{tab:avg}    
\centering        
\begin{tabular}{lcc}
\hline\hline   
 & SN II & SN Ia \\
\hline
\multicolumn{3}{c}{Parameters}\\
\hline
H$\alpha$EW [\AA]                                  & 53.05 {\tiny(14.29)} &28.10 {\tiny(11.30)} \\ 
Age [Myr]                                          &  7.23 {\tiny (0.48)} & 8.72 {\tiny (1.62)} \\ 
log(SFR) [M$_{\sun}$ yr$^{-1}$]                    & -2.64 {\tiny (0.47)} &-2.87 {\tiny (0.18)} \\ 
log($\Sigma$SFR) [M$_{\sun}$ yr$^{-1}$ kpc$^{-2}$] & -1.48 {\tiny (0.45)} &-1.85 {\tiny (0.36)} \\
log(sSFR) [yr$^{-1}$]                              & -9.26 {\tiny (0.15)} &-9.66 {\tiny (0.42)} \\ 
12+log(O/H)$_{\rm M13,O3N2}$  [dex]                &  8.50 {\tiny (0.04)} & 8.55 {\tiny (0.05)} \\ 
\hline
\multicolumn{3}{c}{Cumulative distributions position}\\
\hline
CD H$\alpha$EW                  & 0.59 {\tiny (0.08)} & 0.40 {\tiny (0.14)} \\ 
CD log(SFR)                     & 0.61 {\tiny (0.14)} & 0.61 {\tiny (0.16)} \\ 
CD log($\Sigma$SFR)             & 0.61 {\tiny (0.14)} & 0.63 {\tiny (0.15)} \\ 
CD log(sSFR)                    & 0.49 {\tiny (0.08)} & 0.52 {\tiny (0.17)} \\ 
CD 12+log(O/H)$_{\rm M13,O3N2}$ & 0.57 {\tiny (0.14)} & 0.66 {\tiny (0.22)} \\ 
\hline
\end{tabular}
\end{table}

All local SFR results for all SNe are lower than the nearest nebular cluster measurements, but for SN 2000do. This is expected since the aperture from the SN position include spaxels that are outside of any {\sc Hii} region, and that have supposedly lower H$\alpha$ emission. For SN 2000do in the eastern half of NGC 6754, the extracted aperture was smaller than the size of the neaest {\sc Hii} region, and this explains the different behavior (although note that the difference is compatible with zero).
Note that the behavior of the SFR could be different that the H$\alpha$EW. While the SFR is proportional to the H$\alpha$ intensity, the H$\alpha$EW on the other hand measures how strong the line is compared with the continuum. 
The continuum light is dominated by old stars, which also contain most of the galaxy stellar mass, so H$\alpha$EW can be thought of as an indicator of the strength of the ongoing SFR compared with the past SFR. 

The three SN in NGC 6754 for which aperture spectra $>$2 arcsec were extracted, the $\Sigma$SFR is significantly lower than the value at the nearest {\sc Hii} region (which typically has a diameter $<$2 arcsec). 
Checking the segregation maps, we confirmed that such large apertures include regions of diffuse gas with weak H$\alpha$ emission and thus reducing the $\Sigma$SFR measurement.
Excluding these 3 SN, we find that for 6 out of 8 SNe the $\Sigma$SFR increases in the local spectra with respect the nearest {\sc Hii} region, while two of them have lower values. Although the averages are shifted to lower values, these are biased by the three outliers, and the average difference of the remaining 8 SNe is compatible to zero.

The sSFR is in general higher in the {\sc Hii} region than at SN locations, with the exception of 2 SNe Ia. This stresses that at the locations of SN II, young stars contribute more importantly to the mass content.

No measurements of the local oxygen abundance are available for SN 2010jr in ESO 362-18 and SN 2000cb in IC 1158, since the {\sc [Oiii]} needed is not strong enough to be used reliably.
For the remaining sample, interestingly, we find that while the oxygen abundance at CCSN locations tend to be slightly higher than the abundance of the parent {\sc Hii} region, for SN Ia we find on average lower abundances at SN locations.


\section{Discussion and conclusions} \label{sec:disc}

MUSE observations combine spectral measurements of a wide-field coverage with high spatial resolution, opening a new window for studying SN environments. Observations of NGC 6754 were proposed for Science Verification of the instrument, and after a careful search in other commissioning and SV data, we put together a sample of 6 nearby galaxies that hosted 11 SNe (7 type II, 4 type Ia). Similar analysis was performed on MUSE datacubes as that in  \cite{2012A&A...545A..58S} and \cite{2014A&A...572A..38G}, which previously analysed SN environments within the CALIFA survey. In this analysis we concentrated on extracting all possible {\sc Hii} regions within host galaxies and then characterised the SN host {\sc Hii} region within the context of all other {\sc Hii} regions within each galaxy. To our knowledge, this is the first time that such an approach has been used for the study of SN host galaxies. We extract and analyse {\sc Hii} region statistics of H$\alpha$ EWs, SFRs and $\Sigma$SFRs, together with oxygen abundances. Explosion site properties are hence positioned within the full galaxy-wide distributions of these age and metallicity parameters.

The low numbers of SNe included in this exploratory work do not allow to make any concrete conclusions on SN progenitor properties, however the novel techniques used and presented in e.g. Figs \ref{fig:hiiew}, \ref{fig:hiisfr} and \ref{fig:hiioh} motivate further work on large samples (such as those which will be enabled by the AMUSING project). This work builds on the initial IFS SN environment studies of 
\cite{2014A&A...572A..38G}, and the smaller FOV work by \cite{2013AJ....146...30K, 2013AJ....146...31K},
and shows the possibilities of such data for further understanding SN host galaxies and using these to probe differences in SN progenitors. In conclusion, MUSE promises to be a powerful instrument in the field of SN environment studies for years to come.


\section*{Acknowledgments}

We would like to thank the anonymous referee for his thoughtful comments, and the MUSE Science Verification team for observing the data analyzed in this work, and the PI's of the respective SV proposals.
Support for LG, HK, MH, SGG and FF is provided by the Ministry of Economy, Development, and Tourism's Millennium Science Initiative through grant IC120009, awarded to The Millennium Institute of Astrophysics, MAS. 
LG, HK and FF acknowledge support by CONICYT through FONDECYT grants 3140566, 3140563, and 11130228, respectively.
EP acknowledges support from Spanish MINECO project AYA2014-57490P and Junta de Andaluc\'ia FQ1580.
Based on observations made with ESO Telescopes at the La Silla Paranal Observatory under programs 60.A-9000(B), 60.A-9000(A), 60.A-9329(A), 60.A-9301(A), 60.A-9319(A), 60.A-9339(A).
The data reduction and processing required for the analysis presented in this paper has been computed in a last generation research seed unit server provided by Intel Corporation through the research agreement with the Center for Mathematical Modeling of the University of Chile. 
The authors acknowledge the benefits and the speed of data processing provided by such a cutting edge technology.

\bibliographystyle{MUSE_MNRAS}
\bibliography{biblio}

\begin{thebibliography}{65}
\expandafter\ifx\csname natexlab\endcsname\relax\def\natexlab#1{#1}\fi

\bibitem[{{Alberdi} {et~al}\mbox{.}(2006){Alberdi}, {Colina}, {Torrelles},
  {Panagia}, {Wilson}, \& {Garrington}}]{2006ApJ...638..938A}
{Alberdi} A., {Colina} L., {Torrelles} J.~M., {Panagia} N., {Wilson} A.~S.,
  {Garrington} S.~T., 2006, \apj, 638, 938

\bibitem[{{Alloin} {et~al}\mbox{.}(1979){Alloin}, {Collin-Souffrin}, {Joly}, \&
  {Vigroux}}]{1979A&A....78..200A}
{Alloin} D., {Collin-Souffrin} S., {Joly} M., {Vigroux} L., 1979, \aap, 78, 200

\bibitem[{{Anderson} {et~al}\mbox{.}(2010){Anderson}, {Covarrubias}, {James},
  {Hamuy}, \& {Habergham}}]{2010MNRAS.407.2660A}
{Anderson} J.~P., {Covarrubias} R.~A., {James} P.~A., {Hamuy} M., {Habergham}
  S.~M., 2010, \mnras, 407, 2660

\bibitem[{{Anderson} {et~al}\mbox{.}(2012){Anderson}, {Habergham}, {James}, \&
  {Hamuy}}]{2012MNRAS.424.1372A}
{Anderson} J.~P., {Habergham} S.~M., {James} P.~A., {Hamuy} M., 2012, \mnras,
  424, 1372

\bibitem[{{Anderson} {et~al}\mbox{.}(2015{\natexlab{a}}){Anderson}, {James},
  {F{\"o}rster}, {Gonz{\'a}lez-Gait{\'a}n}, {Habergham}, {Hamuy}, \&
  {Lyman}}]{2015MNRAS.448..732A}
{Anderson} J.~P., {James} P.~A., {F{\"o}rster} F., {Gonz{\'a}lez-Gait{\'a}n}
  S., {Habergham} S.~M., {Hamuy} M., {Lyman} J.~D., 2015{\natexlab{a}}, \mnras,
  448, 732

\bibitem[{{Anderson} {et~al}\mbox{.}(2015{\natexlab{b}}){Anderson}, {James},
  {Habergham}, {Galbany}, \& {Kuncarayakti}}]{2015PASA...32...19A}
{Anderson} J.~P., {James} P.~A., {Habergham} S.~M., {Galbany} L.,
  {Kuncarayakti} H., 2015{\natexlab{b}}, \pasa, 32, 19

\bibitem[{{Bacon} {et~al}\mbox{.}(2014){Bacon}, {Vernet}, {Borisova},
  {Bouch{\'e}}, {Brinchmann}, {Carollo}, {Carton}, {Caruana}, {Cerda},
  {Contini}, {Franx}, {Girard}, {Guerou}, {Haddad}, {Hau}, {Herenz}, {Herrera},
  {Husemann}, {Husser}, {Jarno}, {Kamann}, {Krajnovic}, {Lilly}, {Mainieri},
  {Martinsson}, {Palsa}, {Patricio}, {P{\'e}contal}, {Pello}, {Piqueras},
  {Richard}, {Sandin}, {Schroetter}, {Selman}, {Shirazi}, {Smette}, {Soto},
  {Streicher}, {Urrutia}, {Weilbacher}, {Wisotzki}, \&
  {Zins}}]{2014Msngr.157...13B}
{Bacon} R. {et~al.}, 2014, The Messenger, 157, 13

\bibitem[{{Baldwin}, {Phillips} \& {Terlevich}(1981){Baldwin}, {Phillips}, \&
  {Terlevich}}]{1981PASP...93....5B}
{Baldwin} J.~A., {Phillips} M.~M., {Terlevich} R., 1981, \pasp, 93, 5

\bibitem[{{Boissier} \& {Prantzos}(2009)}]{2009A&A...503..137B}
{Boissier} S., {Prantzos} N., 2009, \aap, 503, 137

\bibitem[{{Bruzual}(2007)}]{2007ASPC..374..303B}
{Bruzual} G., 2007, in Astronomical Society of the Pacific Conference Series,
  Vol. 374, From Stars to Galaxies: Building the Pieces to Build Up the
  Universe, {Vallenari} A., {Tantalo} R., {Portinari} L., {Moretti} A., eds.,
  p. 303

\bibitem[{{Bruzual} \& {Charlot}(2003)}]{2003MNRAS.344.1000B}
{Bruzual} G., {Charlot} S., 2003, \mnras, 344, 1000

\bibitem[{{Castellanos}, {D{\'{\i}}az} \& {Tenorio-Tagle}(2002){Castellanos},
  {D{\'{\i}}az}, \& {Tenorio-Tagle}}]{2002ApJ...565L..79C}
{Castellanos} M., {D{\'{\i}}az} {\'A}.~I., {Tenorio-Tagle} G., 2002, \apjl,
  565, L79

\bibitem[{{Catal{\'a}n-Torrecilla}
  {et~al}\mbox{.}(2015){Catal{\'a}n-Torrecilla}, {Gil de Paz},
  {Castillo-Morales}, {Iglesias-P{\'a}ramo}, {S{\'a}nchez}, {Kennicutt},
  {P{\'e}rez-Gonz{\'a}lez}, {Marino}, {Walcher}, {Husemann},
  {Garc{\'{\i}}a-Benito}, {Mast}, {Gonz{\'a}lez Delgado}, {Mu{\~n}oz-Mateos},
  {Bland-Hawthorn}, {Bomans}, {del Olmo}, {Galbany}, {Gomes}, {Kehrig},
  {L{\'o}pez-S{\'a}nchez}, {Mendoza}, {Monreal-Ibero}, {P{\'e}rez-Torres},
  {S{\'a}nchez-Bl{\'a}zquez}, {Vilchez}, \& {the CALIFA
  collaboration}}]{2015arXiv150703801C}
{Catal{\'a}n-Torrecilla} C. {et~al.}, 2015, ArXiv e-prints

\bibitem[{{Chabrier}(2003)}]{2003PASP..115..763C}
{Chabrier} G., 2003, \pasp, 115, 763

\bibitem[{{Cid Fernandes} {et~al}\mbox{.}(2005){Cid Fernandes}, {Mateus},
  {Sodr{\'e}}, {Stasi{\'n}ska}, \& {Gomes}}]{2005MNRAS.358..363C}
{Cid Fernandes} R., {Mateus} A., {Sodr{\'e}} L., {Stasi{\'n}ska} G., {Gomes}
  J.~M., 2005, \mnras, 358, 363

\bibitem[{{Cid Fernandes} {et~al}\mbox{.}(2009){Cid Fernandes}, {Schoenell},
  {Gomes}, {Asari}, {Schlickmann}, {Mateus}, {Stasinska}, {Sodr{\'e}},
  {Torres-Papaqui}, \& {Seagal Collaboration}}]{2009RMxAC..35..127C}
{Cid Fernandes} R. {et~al.}, 2009, in Revista Mexicana de Astronomia y
  Astrofisica Conference Series, Vol.~35, Revista Mexicana de Astronomia y
  Astrofisica Conference Series, pp. 127--132

\bibitem[{{Filippenko} \& {Matheson}(1993)}]{1993IAUC.5842....2F}
{Filippenko} A.~V., {Matheson} T., 1993, \iaucirc, 5842, 2

\bibitem[{{Fitzpatrick}(1999)}]{1999PASP..111...63F}
{Fitzpatrick} E.~L., 1999, \pasp, 111, 63

\bibitem[{{Freudling} {et~al}\mbox{.}(2013){Freudling}, {Romaniello},
  {Bramich}, {Ballester}, {Forchi}, {Garc{\'{\i}}a-Dabl{\'o}}, {Moehler}, \&
  {Neeser}}]{2013A&A...559A..96F}
{Freudling} W., {Romaniello} M., {Bramich} D.~M., {Ballester} P., {Forchi} V.,
  {Garc{\'{\i}}a-Dabl{\'o}} C.~E., {Moehler} S., {Neeser} M.~J., 2013, \aap,
  559, A96

\bibitem[{{Galbany} {et~al}\mbox{.}(2012){Galbany}, {Miquel}, {{\"O}stman},
  {Brown}, {Cinabro}, {D'Andrea}, {Frieman}, {Jha}, {Marriner}, {Nichol},
  {Nordin}, {Olmstead}, {Sako}, {Schneider}, {Smith}, {Sollerman}, {Pan},
  {Snedden}, {Bizyaev}, {Brewington}, {Malanushenko}, {Malanushenko},
  {Oravetz}, {Simmons}, \& {Shelden}}]{2012ApJ...755..125G}
{Galbany} L. {et~al.}, 2012, \apj, 755, 125

\bibitem[{{Galbany} {et~al}\mbox{.}(2014){Galbany}, {Stanishev}, {Mour{\~a}o},
  {Rodrigues}, {Flores}, {Garc{\'{\i}}a-Benito}, {Mast}, {Mendoza},
  {S{\'a}nchez}, {Badenes}, {Barrera-Ballesteros}, {Bland-Hawthorn},
  {Falc{\'o}n-Barroso}, {Garc{\'{\i}}a-Lorenzo}, {Gomes}, {Gonz{\'a}lez
  Delgado}, {Kehrig}, {Lyubenova}, {L{\'o}pez-S{\'a}nchez}, {de
  Lorenzo-C{\'a}ceres}, {Marino}, {Meidt}, {Moll{\'a}}, {Papaderos},
  {P{\'e}rez-Torres}, {Rosales-Ortega}, \& {van de Ven}}]{2014A&A...572A..38G}
{Galbany} L. {et~al.}, 2014, \aap, 572, A38

\bibitem[{{Giammanco}, {Beckman} \& {Cedr{\'e}s}(2005){Giammanco}, {Beckman},
  \& {Cedr{\'e}s}}]{2005A&A...438..599G}
{Giammanco} C., {Beckman} J.~E., {Cedr{\'e}s} B., 2005, \aap, 438, 599

\bibitem[{{Gonz\'alez Delgado} \& {P\'erez}(1997)}]{1997ApJS..108..199G}
{Gonz\'alez Delgado} R.~M., {P\'erez} E., 1997, \apjs, 108, 199

\bibitem[{{Gonz{\'a}lez-Gait{\'a}n}
  {et~al}\mbox{.}(2015){Gonz{\'a}lez-Gait{\'a}n}, {Tominaga}, {Molina},
  {Galbany}, {Bufano}, {Anderson}, {Gutierrez}, {F{\"o}rster}, {Pignata},
  {Bersten}, {Howell}, {Sullivan}, {Carlberg}, {de Jaeger}, {Hamuy},
  {Baklanov}, \& {Blinnikov}}]{2015MNRAS.451.2212G}
{Gonz{\'a}lez-Gait{\'a}n} S. {et~al.}, 2015, \mnras, 451, 2212

\bibitem[{{Hamuy}(2003)}]{2003ApJ...582..905H}
{Hamuy} M., 2003, \apj, 582, 905

\bibitem[{{Izotov} {et~al}\mbox{.}(2006){Izotov}, {Stasi{\'n}ska}, {Meynet},
  {Guseva}, \& {Thuan}}]{2006A&A...448..955I}
{Izotov} Y.~I., {Stasi{\'n}ska} G., {Meynet} G., {Guseva} N.~G., {Thuan} T.~X.,
  2006, \aap, 448, 955

\bibitem[{{Kauffmann} {et~al}\mbox{.}(2003){Kauffmann}, {Heckman}, {Tremonti},
  {Brinchmann}, {Charlot}, {White}, {Ridgway}, {Brinkmann}, {Fukugita}, {Hall},
  {Ivezi{\'c}}, {Richards}, \& {Schneider}}]{2003MNRAS.346.1055K}
{Kauffmann} G. {et~al.}, 2003, \mnras, 346, 1055

\bibitem[{{Kelly} {et~al}\mbox{.}(2015){Kelly}, {Filippenko}, {Burke},
  {Hicken}, {Ganeshalingam}, \& {Zheng}}]{2015Sci...347.1459K}
{Kelly} P.~L., {Filippenko} A.~V., {Burke} D.~L., {Hicken} M., {Ganeshalingam}
  M., {Zheng} W., 2015, Science, 347, 1459

\bibitem[{{Kelly} \& {Kirshner}(2012)}]{2012ApJ...759..107K}
{Kelly} P.~L., {Kirshner} R.~P., 2012, \apj, 759, 107

\bibitem[{{Kelly}, {Kirshner} \& {Pahre}(2008){Kelly}, {Kirshner}, \&
  {Pahre}}]{2008ApJ...687.1201K}
{Kelly} P.~L., {Kirshner} R.~P., {Pahre} M., 2008, \apj, 687, 1201

\bibitem[{{Kennicutt}(1998)}]{1998ApJ...498..541K}
{Kennicutt}, Jr. R.~C., 1998, \apj, 498, 541

\bibitem[{{Kewley} {et~al}\mbox{.}(2001){Kewley}, {Dopita}, {Sutherland},
  {Heisler}, \& {Trevena}}]{2001ApJ...556..121K}
{Kewley} L.~J., {Dopita} M.~A., {Sutherland} R.~S., {Heisler} C.~A., {Trevena}
  J., 2001, \apj, 556, 121

\bibitem[{{Kuncarayakti} {et~al}\mbox{.}(2013{\natexlab{a}}){Kuncarayakti},
  {Doi}, {Aldering}, {Arimoto}, {Maeda}, {Morokuma}, {Pereira}, {Usuda}, \&
  {Hashiba}}]{2013AJ....146...30K}
{Kuncarayakti} H. {et~al.}, 2013{\natexlab{a}}, \aj, 146, 30

\bibitem[{{Kuncarayakti} {et~al}\mbox{.}(2013{\natexlab{b}}){Kuncarayakti},
  {Doi}, {Aldering}, {Arimoto}, {Maeda}, {Morokuma}, {Pereira}, {Usuda}, \&
  {Hashiba}}]{2013AJ....146...31K}
{Kuncarayakti} H. {et~al.}, 2013{\natexlab{b}}, \aj, 146, 31

\bibitem[{{Leitherer} {et~al}\mbox{.}(1999){Leitherer}, {Schaerer}, {Goldader},
  {Delgado}, {Robert}, {Kune}, {de Mello}, {Devost}, \&
  {Heckman}}]{1999ApJS..123....3L}
{Leitherer} C. {et~al.}, 1999, \apjs, 123, 3

\bibitem[{{Leloudas} {et~al}\mbox{.}(2011){Leloudas}, {Gallazzi}, {Sollerman},
  {Stritzinger}, {Fynbo}, {Hjorth}, {Malesani}, {Micha{\l}owski},
  {Milvang-Jensen}, \& {Smith}}]{2011A&A...530A..95L}
{Leloudas} G. {et~al.}, 2011, \aap, 530, A95

\bibitem[{{L\'opez} {et~al}\mbox{.}(2011){L\'opez}, {Krumholz}, {Bolatto},
  {Prochaska}, \& {Ramirez-Ruiz}}]{2011ApJ...731...91L}
{L\'opez} L.~A., {Krumholz} M.~R., {Bolatto} A.~D., {Prochaska} J.~X.,
  {Ramirez-Ruiz} E., 2011, \apj, 731, 91

\bibitem[{{L{\'o}pez-S{\'a}nchez} \& {Esteban}(2010)}]{2010A&A...517A..85L}
{L{\'o}pez-S{\'a}nchez} {\'A}.~R., {Esteban} C., 2010, \aap, 517, A85

\bibitem[{{Marigo} \& {Girardi}(2007)}]{2007A&A...469..239M}
{Marigo} P., {Girardi} L., 2007, \aap, 469, 239

\bibitem[{{Marigo} {et~al}\mbox{.}(2008){Marigo}, {Girardi}, {Bressan},
  {Groenewegen}, {Silva}, \& {Granato}}]{2008A&A...482..883M}
{Marigo} P., {Girardi} L., {Bressan} A., {Groenewegen} M.~A.~T., {Silva} L.,
  {Granato} G.~L., 2008, \aap, 482, 883

\bibitem[{{Marino} {et~al}\mbox{.}(2013){Marino}, {Rosales-Ortega},
  {S{\'a}nchez}, {Gil de Paz}, {V{\'{\i}}lchez}, {Miralles-Caballero},
  {Kehrig}, {P{\'e}rez-Montero}, {Stanishev}, {Iglesias-P{\'a}ramo},
  {D{\'{\i}}az}, {Castillo-Morales}, {Kennicutt}, {L{\'o}pez-S{\'a}nchez},
  {Galbany}, {Garc{\'{\i}}a-Benito}, {Mast}, {Mendez-Abreu}, {Monreal-Ibero},
  {Husemann}, {Walcher}, {Garc{\'{\i}}a-Lorenzo}, {Masegosa}, {Del Olmo
  Orozco}, {Mour{\~a}o}, {Ziegler}, {Moll{\'a}}, {Papaderos},
  {S{\'a}nchez-Bl{\'a}zquez}, {Gonz{\'a}lez Delgado}, {Falc{\'o}n-Barroso},
  {Roth}, {van de Ven}, \& {Califa Team}}]{2013A&A...559A.114M}
{Marino} R.~A. {et~al.}, 2013, \aap, 559, A114

\bibitem[{{Mast} {et~al}\mbox{.}(2014){Mast}, {Rosales-Ortega}, {S{\'a}nchez},
  {V{\'{\i}}lchez}, {Iglesias-Paramo}, {Walcher}, {Husemann}, {M{\'a}rquez},
  {Marino}, {Kennicutt}, {Monreal-Ibero}, {Galbany}, {de Lorenzo-C{\'a}ceres},
  {Mendez-Abreu}, {Kehrig}, {del Olmo}, {Rela{\~n}o}, {Wisotzki},
  {M{\'a}rmol-Queralt{\'o}}, {Bekerait{\`e}}, {Papaderos}, {Wild}, {Aguerri},
  {Falc{\'o}n-Barroso}, {Bomans}, {Ziegler}, {Garc{\'{\i}}a-Lorenzo},
  {Bland-Hawthorn}, {L{\'o}pez-S{\'a}nchez}, \& {van de
  Ven}}]{2014A&A...561A.129M}
{Mast} D. {et~al.}, 2014, \aap, 561, A129

\bibitem[{{Mattila} {et~al}\mbox{.}(2005){Mattila}, {Lundqvist}, {Sollerman},
  {Kozma}, {Baron}, {Fransson}, {Leibundgut}, \&
  {Nomoto}}]{2005A&A...443..649M}
{Mattila} S., {Lundqvist} P., {Sollerman} J., {Kozma} C., {Baron} E.,
  {Fransson} C., {Leibundgut} B., {Nomoto} K., 2005, \aap, 443, 649

\bibitem[{{Modjaz} {et~al}\mbox{.}(2011){Modjaz}, {Kewley}, {Bloom},
  {Filippenko}, {Perley}, \& {Silverman}}]{2011ApJ...731L...4M}
{Modjaz} M., {Kewley} L., {Bloom} J.~S., {Filippenko} A.~V., {Perley} D.,
  {Silverman} J.~M., 2011, \apjl, 731, L4

\bibitem[{{Moustakas} {et~al}\mbox{.}(2010){Moustakas}, {Kennicutt},
  {Tremonti}, {Dale}, {Smith}, \& {Calzetti}}]{2010ApJS..190..233M}
{Moustakas} J., {Kennicutt}, Jr. R.~C., {Tremonti} C.~A., {Dale} D.~A., {Smith}
  J.-D.~T., {Calzetti} D., 2010, ApJS, 190, 233

\bibitem[{{Osterbrock} \& {Ferland}(2006)}]{2006agna.book.....O}
{Osterbrock} D.~E., {Ferland} G.~J., 2006, {Astrophysics of gaseous nebulae and
  active galactic nuclei}

\bibitem[{{Pastorello} {et~al}\mbox{.}(2005){Pastorello}, {Baron}, {Branch},
  {Zampieri}, {Turatto}, {Ramina}, {Benetti}, {Cappellaro}, {Salvo}, {Patat},
  {Piemonte}, {Sollerman}, {Leibundgut}, \& {Altavilla}}]{2005MNRAS.360..950P}
{Pastorello} A. {et~al.}, 2005, \mnras, 360, 950

\bibitem[{{Peimbert} {et~al}\mbox{.}(2007){Peimbert}, {Peimbert}, {Esteban},
  {Garc{\'{\i}}a-Rojas}, {Bresolin}, {Carigi}, {Ruiz}, \&
  {L{\'o}pez-S{\'a}nchez}}]{2007RMxAC..29...72P}
{Peimbert} M., {Peimbert} A., {Esteban} C., {Garc{\'{\i}}a-Rojas} J.,
  {Bresolin} F., {Carigi} L., {Ruiz} M.~T., {L{\'o}pez-S{\'a}nchez} A.~R.,
  2007, in Revista Mexicana de Astronomia y Astrofisica Conference Series,
  Vol.~29, Revista Mexicana de Astronomia y Astrofisica Conference Series,
  {Guzm{\'a}n} R., ed., pp. 72--79

\bibitem[{{Prieto}, {Stanek} \& {Beacom}(2008){Prieto}, {Stanek}, \&
  {Beacom}}]{2008ApJ...673..999P}
{Prieto} J.~L., {Stanek} K.~Z., {Beacom} J.~F., 2008, \apj, 673, 999

\bibitem[{{Rigault} {et~al}\mbox{.}(2013){Rigault}, {Copin}, {Aldering},
  {Antilogus}, {Aragon}, {Bailey}, {Baltay}, {Bongard}, {Buton}, {Canto},
  {Cellier-Holzem}, {Childress}, {Chotard}, {Fakhouri}, {Feindt}, {Fleury},
  {Gangler}, {Greskovic}, {Guy}, {Kim}, {Kowalski}, {Lombardo}, {Nordin},
  {Nugent}, {Pain}, {P{\'e}contal}, {Pereira}, {Perlmutter}, {Rabinowitz},
  {Runge}, {Saunders}, {Scalzo}, {Smadja}, {Tao}, {Thomas}, \&
  {Weaver}}]{2013A&A...560A..66R}
{Rigault} M. {et~al.}, 2013, \aap, 560, A66

\bibitem[{{Ruiz-Lapuente}(2014)}]{2014NewAR..62...15R}
{Ruiz-Lapuente} P., 2014, \nar, 62, 15

\bibitem[{{S{\'a}nchez} {et~al}\mbox{.}(2015){S{\'a}nchez}, {Galbany},
  {P{\'e}rez}, {S{\'a}nchez-Bl{\'a}zquez}, {Falc{\'o}n-Barroso},
  {Rosales-Ortega}, {S{\'a}nchez-Menguiano}, {Marino}, {Kuncarayakti},
  {Anderson}, {Kruehler}, {Cano-D{\'{\i}}az}, {Barrera-Ballesteros}, \&
  {Gonz{\'a}lez-Gonz{\'a}lez}}]{2015A&A...573A.105S}
{S{\'a}nchez} S.~F. {et~al.}, 2015, \aap, 573, A105

\bibitem[{{S{\'a}nchez} {et~al}\mbox{.}(2014){S{\'a}nchez}, {Rosales-Ortega},
  {Iglesias-P{\'a}ramo}, {Moll{\'a}}, {Barrera-Ballesteros}, {Marino},
  {P{\'e}rez}, {S{\'a}nchez-Blazquez}, {Gonz{\'a}lez Delgado}, {Cid Fernandes},
  {de Lorenzo-C{\'a}ceres}, {Mendez-Abreu}, {Galbany}, {Falcon-Barroso},
  {Miralles-Caballero}, {Husemann}, {Garc{\'{\i}}a-Benito}, {Mast}, {Walcher},
  {Gil de Paz}, {Garc{\'{\i}}a-Lorenzo}, {Jungwiert}, {V{\'{\i}}lchez},
  {J{\'{\i}}lkov{\'a}}, {Lyubenova}, {Cortijo-Ferrero}, {D{\'{\i}}az},
  {Wisotzki}, {M{\'a}rquez}, {Bland-Hawthorn}, {Ellis}, {van de Ven}, {Jahnke},
  {Papaderos}, {Gomes}, {Mendoza}, \&
  {L{\'o}pez-S{\'a}nchez}}]{2014A&A...563A..49S}
{S{\'a}nchez} S.~F. {et~al.}, 2014, \aap, 563, A49

\bibitem[{{S{\'a}nchez} {et~al}\mbox{.}(2012){S{\'a}nchez}, {Rosales-Ortega},
  {Marino}, {Iglesias-P{\'a}ramo}, {V{\'{\i}}lchez}, {Kennicutt},
  {D{\'{\i}}az}, {Mast}, {Monreal-Ibero}, {Garc{\'{\i}}a-Benito},
  {Bland-Hawthorn}, {P{\'e}rez}, {Gonz{\'a}lez Delgado}, {Husemann},
  {L{\'o}pez-S{\'a}nchez}, {Cid Fernandes}, {Kehrig}, {Walcher}, {Gil de Paz},
  \& {Ellis}}]{2012A&A...546A...2S}
{S{\'a}nchez} S.~F. {et~al.}, 2012, \aap, 546, A2

\bibitem[{{S{\'a}nchez-Bl{\'a}zquez}
  {et~al}\mbox{.}(2006){S{\'a}nchez-Bl{\'a}zquez}, {Peletier},
  {Jim{\'e}nez-Vicente}, {Cardiel}, {Cenarro}, {Falc{\'o}n-Barroso}, {Gorgas},
  {Selam}, \& {Vazdekis}}]{2006MNRAS.371..703S}
{S{\'a}nchez-Bl{\'a}zquez} P. {et~al.}, 2006, \mnras, 371, 703

\bibitem[{{Sanders} {et~al}\mbox{.}(2012){Sanders}, {Soderberg}, {Levesque},
  {Foley}, {Chornock}, {Milisavljevic}, {Margutti}, {Berger}, {Drout},
  {Czekala}, \& {Dittmann}}]{2012ApJ...758..132S}
{Sanders} N.~E. {et~al.}, 2012, \apj, 758, 132

\bibitem[{{Schlafly} \& {Finkbeiner}(2011)}]{2011ApJ...737..103S}
{Schlafly} E.~F., {Finkbeiner} D.~P., 2011, \apj, 737, 103

\bibitem[{{Shao} {et~al}\mbox{.}(2014){Shao}, {Liang}, {Dennefeld}, {Chen},
  {Zhong}, {Hammer}, {Deng}, {Flores}, {Zhang}, {Shi}, \&
  {Zhou}}]{2014ApJ...791...57S}
{Shao} X. {et~al.}, 2014, \apj, 791, 57

\bibitem[{{Stanishev} {et~al}\mbox{.}(2012){Stanishev}, {Rodrigues},
  {Mour{\~a}o}, \& {Flores}}]{2012A&A...545A..58S}
{Stanishev} V., {Rodrigues} M., {Mour{\~a}o} A., {Flores} H., 2012, \aap, 545,
  A58

\bibitem[{{Stasi{\'n}ska}(2006)}]{2006A&A...454L.127S}
{Stasi{\'n}ska} G., 2006, \aap, 454, L127

\bibitem[{{Stritzinger} {et~al}\mbox{.}(2012){Stritzinger}, {Taddia},
  {Fransson}, {Fox}, {Morrell}, {Phillips}, {Sollerman}, {Anderson}, {Boldt},
  {Brown}, {Campillay}, {Castellon}, {Contreras}, {Folatelli}, {Habergham},
  {Hamuy}, {Hjorth}, {James}, {Krzeminski}, {Mattila}, {Persson}, \&
  {Roth}}]{2012ApJ...756..173S}
{Stritzinger} M. {et~al.}, 2012, \apj, 756, 173

\bibitem[{{Taddia} {et~al}\mbox{.}(2015){Taddia}, {Sollerman}, {Fremling},
  {Pastorello}, {Leloudas}, {Fransson}, {Nyholm}, {Stritzinger}, {Ergon},
  {Roy}, \& {Migotto}}]{2015A&A...580A.131T}
{Taddia} F. {et~al.}, 2015, \aap, 580, A131

\bibitem[{{Taddia} {et~al}\mbox{.}(2012){Taddia}, {Stritzinger}, {Sollerman},
  {Phillips}, {Anderson}, {Ergon}, {Folatelli}, {Fransson}, {Freedman},
  {Hamuy}, {Morrell}, {Pastorello}, {Persson}, \&
  {Gonzalez}}]{2012A&A...537A.140T}
{Taddia} F. {et~al.}, 2012, \aap, 537, A140

\bibitem[{{Veilleux} \& {Osterbrock}(1987)}]{1987ApJS...63..295V}
{Veilleux} S., {Osterbrock} D.~E., 1987, ApJS, 63, 295

\bibitem[{{Weilbacher} {et~al}\mbox{.}(2014){Weilbacher}, {Streicher},
  {Urrutia}, {P{\'e}contal-Rousset}, {Jarno}, \& {Bacon}}]{2014ASPC..485..451W}
{Weilbacher} P.~M., {Streicher} O., {Urrutia} T., {P{\'e}contal-Rousset} A.,
  {Jarno} A., {Bacon} R., 2014, in Astronomical Society of the Pacific
  Conference Series, Vol. 485, Astronomical Data Analysis Software and Systems
  XXIII, {Manset} N., {Forshay} P., eds., p. 451

\end{thebibliography}

\end{document}